\documentclass{aa}  
\usepackage{graphicx}
\usepackage{natbib}
\usepackage{appendix}
\usepackage{latexsym}
\usepackage{amssymb}

\begin{document}

\title{The formation of ethylene glycol and other complex organic molecules in star-forming regions}

\author{V. M. Rivilla\inst{1}, M. T. Beltr\'an\inst{1}, R. Cesaroni\inst{1}, F. Fontani\inst{1}, C. Codella\inst{1} \and Q. Zhang\inst{2}
}

\institute{Osservatorio Astrofisico di Arcetri, Largo Enrico Fermi 5, I-50125, Firenze, Italy \\
\email{rivilla@arcetri.astro.it}
\and
Harvard-Smithsonian Center for Astrophysics, 60 Garden St., Cambridge, MA 02138, USA \\
}

\date{Received; accepted}

\titlerunning{The formation of ethylene glycol and other COMs in star-forming regions}
\authorrunning{Rivilla et al.}
 
\abstract
{The detection of complex organic molecules related with prebiotic chemistry in star-forming regions allows us to investigate how the basic building blocks of life are formed.}
{Ethylene glycol (CH$_{2}$OH)$_{2}$ is the simplest sugar alcohol, and the reduced alcohol of the simplest sugar glycoladehyde (CH$_{2}$OHCHO). We aim to study the molecular abundance and spatial distribution of (CH$_{2}$OH)$_{2}$, CH$_{2}$OHCHO and other chemically related complex organic species (CH$_{3}$OCHO, CH$_{3}$OCH$_{3}$, and C$_{2}$H$_{5}$OH) towards the chemically rich massive star-forming region G31.41+0.31.}
{We have analyzed multiple single dish (Green Bank Telescope and IRAM 30m) and interferometric (Submillimeter Array) spectra towards G31.41+0.31, covering a range of frequencies from 45 to 258 GHz. We have fitted the observed spectra with a Local Thermodynamic Equilibrium synthetic spectra, and obtained excitation temperatures and column densities. We have compared our findings in G31.41+0.31 with the results found in other environments, including low- and high-mass star-forming regions, cold clouds and comets.}  
{We have reported for the first time the presence of the aGg' conformer of (CH$_{2}$OH)$_{2}$ towards G31.41+0.31, detecting more than 30 unblended lines. We have detected also multiple transitions of other complex organic molecules such as CH$_{2}$OHCHO, CH$_{3}$OCHO, CH$_{3}$OCH$_{3}$ and C$_{2}$H$_{5}$OH. The high angular resolution images show that the (CH$_{2}$OH)$_{2}$ emission is very compact, peaking towards the maximum of the 1.3 mm continuum. These observations suggest that low abundance complex organic molecules, like (CH$_{2}$OH)$_{2}$ or CH$_{2}$OHCHO, are good probes of the gas located closer to the forming stars. Our analysis confirms that (CH$_{2}$OH)$_{2}$ is more abundant than CH$_{2}$OHCHO in G31.41+0.31, as previously observed in other interstellar regions. Comparing different star-forming regions we find evidence of an increase of the (CH$_{2}$OH)$_{2}$/CH$_{2}$OHCHO abundance ratio with the luminosity of the source. The CH$_{3}$OCH$_{3}$/CH$_{3}$OCHO and (CH$_{2}$OH)$_{2}$/C$_{2}$H$_{5}$OH ratios are nearly constant with luminosity. We have also found that the abundance ratios of pairs of isomers (CH$_{2}$OHCHO/CH$_{3}$OCHO and C$_{2}$H$_{5}$OH/CH$_{3}$OCH$_{3}$) decrease with the luminosity of the sources.} 
{The most likely explanation for the behavior of the (CH$_{2}$OH)$_{2}$/CH$_{2}$OHCHO ratio is that these molecules are formed by different chemical formation routes not directly linked; although warm-up timescales effects and different formation and destruction efficiencies in the gas phase cannot be ruled out. The most likely formation route of (CH$_{2}$OH)$_{2}$ is by combination of two CH$_{2}$OH radicals on dust grains. We also favor that CH$_{2}$OHCHO is formed via the solid-phase dimerization of the formyl radical HCO. The interpretation of the observations also suggests a chemical link between CH$_{3}$OCHO and CH$_{3}$OCH$_{3}$.}
  
\keywords{}              

\maketitle

\section{Introduction}

In the past few years, the improvement of the sensitivity of current radio and (sub)millimeter telescopes has allowed us to detect molecular species with increasing number of atoms in the interstellar medium (ISM). From the astrophysical point of view, molecules with $\geq$6 atoms containing carbon are considered as complex organic molecules (COMs). These molecules are expected to play an important role in prebiotic chemistry and are considered the basic ingredients to explain the origin of life. Therefore, understanding how these COMs are formed is a first and unavoidable step to ascertain how life could emerge in the Universe. The detection of COMs in the gas environment where low- and high-mass stars are forming (known as hot corinos and hot cores, respectively) indicates that they are part of the material of which stars, planets and comets are made. Did the Earth inherit the chemical composition from its parental ISM? Were the organic compounds delivered into the early Earth by a rain of meteorites? The answer to these open questions is only possible by studying the chemical compositions of comets and star-forming regions, and figuring out how they were produced.  

The formation of COMs is being intensively debated in astrochemistry. Two possible routes, which can be complementary, have been proposed: gas-phase chemical reactions and reactions on the surface of interstellar dust grains. Only the detection of numerous COMs and the study of their relative abundances and spatial distributions in a large sample of star-forming regions will help us to constrain the chemical pathways leading to the formation of COMs. 

Among the COMs, one finds the basic building blocks of biochemistry: aminoacids (precursors of proteins), monosaccharides (the simplest sugars), and lipids. So far, only the interstellar search for members of the sugar family has been successful. Sugars are a key ingredient in astrobiology because they are associated with both metabolism and genetic information. The simplest sugar, glycolaldehyde (CH$_2$OHCHO, hereafter GA), and the simplest sugar alcohol, ethylene glycol ((CH$_{2}$OH)$_2$, hereafter EG), are among the largest molecules detected so far in the ISM (8 and 10 atoms, respectively). It has been proposed that GA is involved in the formation of aminoacids and more complex sugars. GA can react with another sugar, propenal, to produce ribose, the central constituent of the backbone of ribonucleic acid RNA (\citealt{collins&ferrier95,weber98,krishnamurthy99}). 

EG is the reduced alcohol of GA, and it was first found in the ISM also towards the Galactic Center (\citealt{hollis02,requena-torres08,belloche13}). In the ISM it has been reported towards NGC 1333$-$IRAS 2A (\citealt{maury14}), NGC 7129 FIRS2 (\citealt{fuente14}) and the Orion, W51e1e2 and G34.3+0.1 hot cores (\citealt{brouillet15,lykke15}). Additionally, EG has also been detected in the comets Hale-Bopp (\citealt{crovisier04a}), Lemmon and Lovejoy (\citealt{biver14,biver15}) and 67P/Churyumov-Gerasimenko (\citealt{goesmann15}), and in the meteorites Murchinson and Murray (\citealt{cooper01}).

Theoretical chemical modeling and experimental works have proposed different mechanisms for the formation of EG (and other COMs) in the ISM (\citealt{sorrell01,charnley05,bennett07,garrod08,woods12,woods13,fedoseev15,butscher15}). In order to constrain the most likely chemical pathway, it is essential to confirm the presence of these COMs in more interstellar sources, and to study their spatial distribution and relative abundances. In particular, the massive star-forming regions are a very suitable laboratory to study molecular complexity. The radiation from the forming massive stars warms up the grain mantles of the interstellar dust, triggering the desorption of molecules to the gas phase and enriching the chemical environment. 

In this work we present single-dish and interferometric observations towards the G31.41+0.31 massive star-forming region (hereafter G31). G31, located at a distance of 7.9 kpc (\citealt{churchwell90}), harbors a luminous hot molecular core ($\sim$3$\times$10$^{5}$ L$_{\odot}$, \citealt{beltran15}) believed to be heated by several massive protostars. Many COMs have already been detected towards G31 (see e.g. \citealt{beltran05,fontani07,isokoski13}): methanol (CH$_3$OH), ethanol (C$_2$H$_5$OH), ethyl cyanide (C$_2$H$_5$CN), methyl formate (CH$_3$OCHO), dimethyl ether (CH$_3$OCH$_3$). This chemical richness, along with the previous detection of GA (\citealt{beltran09}), makes G31 an excellent target to search for even more complex species, and EG in particular.

We report on the detection of the aGg' conformer of EG towards G31. We note that this is the first hot core outside the GC where both EG and GA have been detected.
In Section \ref{observations} we present the different observational campaigns. Section \ref{data-analysis} explains the procedures and tools used to identify the lines. The results of the study, including the derivation of physical parameters and spatial distribution of the emission are presented in Section \ref{results}. In Section \ref{comparison} we compare our results in G31 with other interstellar regions and comets. In Section \ref{discussion} we discuss the implications of our findings on the formation pathways of EG and GA and other COMs. Finally, we summarize our conclusions in Section \ref{conclusions}.

%-----------------------------Table Start-----------------------------

\begin{table*}
%\begin{scriptsize}
\caption[]{Summary of observations towards hot molecular core G31 used in this work.}
\begin{center}
%\vspace{-4mm}
\begin{tabular}{c c c c c c}
\hline
Telescope & Date & \multicolumn{2}{c}{Phase center} & Frequency coverage & $\theta_{\rm beam}$    \\
          &      & $RA_{\rm J2000}$ (h m s) &  $DEC_{\rm J2000}$ ($^{\circ}$ $^{\prime}$ $ ^{\prime\prime}$)   & (GHz) & ($\arcsec$)   \\ 
\cline{3-4}
\hline       
IRAM 30m  & 1996 July      & 18 47 34.4 & -01 12 46.0 & 86-258$^{1}$ & 12$-$22 \\ 
          & 2014 Sept. & 18 47 34.3 & -01 12 45.9  & 81.2$-$89.0, 167.93$-$175.71 & 28, 14  \\ 
\hline
GBT & 2011 May & 18 47 34.3 & -01 12 45.8 & 45.14$-$45.92 & 16 \\
\hline
SMA & 2007 May  & 18 47 34.3 & -01 12 45.9  &  219.45$-$221.45, 229.35$-$231.35 & 1.7$\times$ 3.5 (comp.) \\
    & $\&$ July & & &  & 0.90$\times$0.75 (very ext.+ comp.) \\
\hline
\end{tabular}
\end{center}
\vspace{-2mm}
{${(1)}$ This range was not fully covered, only some narrower spectral windows.} \\
%\vspace{-4mm}
\label{table-observations}
%\end{scriptsize}
\end{table*}   

%-------------------

\section{Observations}
\label{observations}

In this work we have used observations performed with the single-dish telescopes GBT and IRAM 30m, and the SMA interferometer, which provide us with both wide spectral coverage and spatial information. Technical details of the observations are given in the following subsections.

\subsection{Green Bank Telescope (GBT) single-dish observations}

The observations were carried out on May 2011 using the Q band receiver (7 mm). We used the maximum bandwidth provided by the GBT spectrometer ($\sim$800 MHz), covering the range 45.14$-$45.92 GHz. To avoid irregular band-pass shapes related to the wide bandwidth spectrometer mode, we used the sub-reflector nodding technique, in which the source is positioned first in one receiver feed and then in the other. Moving the subreflector we minimized the time spent between feeds. The advantage of this technique is that one feed is always observing the source and the calibration can be done more accurately. We observed two overlap spectral windows with dual polarization, with a spectral resolution of 0.39 MHz (2.5 km s$^{-1}$). The system temperature was $\sim$80 K. We applied  out-of-focus (OOF) holography to correct large-scale errors in the shape of the reflecting surface at the beginning of the observing run. The reduction was done with the GBT-IDL package, and the data were exported to {\it MADCUBAIJ}\footnote{Madrid Data Cube Analysis on ImageJ is a software developed in the Center of Astrobiology (Madrid, INTA-CSIC) to visualize and analyze single spectra and datacubes (Mart\'in et al., {\it in prep.}} for the analysis.

The GBT telescope provided the spectra in antenna temperature below the atmosphere, $T_{\rm a}$. We converted this scale to antenna temperature above the atmosphere, T$_{\rm a}^{*}$, using the expression $T_{\rm a}=T_{\rm a}^{*}\times$ exp$(-\tau / m_{\rm air})$, where $\tau$ is the opacity at zenith and $m_{\rm air}$ is the air mass, which was estimated with 1/sin(elevation). Then, T$_{\rm a}^{*}$ was converted to main beam temperature with $T_{\rm mb}=T_{\rm a}^{*}/B_{\rm eff}$, considering a beam efficiency of $B_{\rm eff}$=0.83.

\subsection{IRAM 30m single-dish observations}

We have used data from two different observational campaigns performed with the IRAM 30m telescope at Pico Veleta (Spain) on July 1996 and September 2014. The phase center, frequency coverage and beams of these observations are given in Table \ref{table-observations}. The 1996 observations were carried out simultaneously at 3, 2 and 1.3 mm. 
For more details, see \citet{fontani07}.The data were reduced using the software package CLASS of GILDAS and then exported to {\it MADCUBAIJ} for analysis.

The 2014 observations used the Eight Mixer Receiver (EMIR) and the Fast Fourier Transform Spectrometer (FTS) covering the frequency ranges 81.23$-$89.01 GHz (3 mm) and 167.93$-$175.71-GHz (2 mm). The spectral resolution was 0.195 kHz (0.34 km s$^{-1}$ and 0.67 km s$^{-1}$ for 2 and 3 mm, respectively). The system temperatures were $\sim$130 K and $\sim$900 K for the 3 and 2 mm receivers, respectively.
The spectra were exported from CLASS to {\it MADCUBAIJ}, which was used for the reduction and analysis. To increase the signal to noise in the 2 mm spectra, we smoothed the data to a velocity resolution of 1 km s$^{-1}$. 

The line intensity of the IRAM 30m spectra was converted to the main beam temperature T$_{\rm mb}$, which were calculated as: $T_{\rm mb}$=$T_{\rm a}^{*} {F_{\rm eff}}/{B_{\rm eff}}$, where $T_{\rm a}^{*}$ is the antenna temperature, and $F_{\rm eff}$ and $B_{\rm eff}$ are the forward and beam efficiencies, respectively. For the 2014 observations we used the ratios ${F_{\rm eff}}/{B_{\rm eff}}$ 1.17 and 1.37 for 3 and 2 mm, respectively. The beams of the observations can be estimated using the expression: $\theta_{\rm beam}$[arcsec]=2460/${\nu(\rm GHz)}$.

 %-----------------------------Table Start-----------------------------
\begin{table*}
%\begin{scriptsize}
\caption[]{Summary of the detections of the different COMs in the different telescope datasets.}
\begin{center}
%\vspace{-4mm}
\begin{tabular}{c c c c c c}
\hline
Telescope & (CH$_2$OH)$_2$ (EG) & C$_2$H$_5$OH (ET) & CH$_2$OHCHO (GA)  & CH$_3$OCHO (MF) & DME (CH$_3$OCH$_3$) \\
\hline
GBT     &  $\checkmark$ & $\checkmark$   &  ---$^{a}$ & $\checkmark$ & ---$^{a}$  \\
IRAM 30m       & $\checkmark$  & $\checkmark$   & $\checkmark$  & $\checkmark$ & $\checkmark$   \\
SMA & $\checkmark$ & $\checkmark$  &  $\checkmark$ & $\checkmark$ & $\checkmark$  \\ 
\hline
\end{tabular}
\end{center}
%\vspace{-3mm}
{$^{a}$The not detections are because there are no bright lines of these species in the frequency range covered by the GBT observations.} \\
%{$^{b}$ Integrated intensities derived from the LTE fit.} \\
\label{table-detection-summary}
%\end{scriptsize}
\end{table*}  

%-----------------------------Table Start-----------------------------

\begin{table*}
%\begin{scriptsize}
\caption[]{Clean transitions (i.e. non blended with other molecular species) of EG identified in the single dish spectra (GBT and IRAM 30m) towards G31.}
\begin{center}
%\vspace{-4mm}
\begin{tabular}{c c c c c}
\hline
Frequency &  Transition &  $E_{\rm up}$  & $\int$ T$_{\rm mb}\times\Delta v$  & Panel \\
(GHz) &  $J_{K_a,K_c}, v$  & (K) & (K$\times$km s$^{-1}$) & Fig. \ref{fig-single-clean}\\
\hline\hline
45.17972       &   5$_{2,3}$ (v=0) $-$   4$_{2,2}$ (v=1) &   9 & 0.351 & a) \\
45.54764       &   4$_{1,4}$ (v=1) $-$   3$_{1,3}$ (v=0) &   5 & 0.305 & b) \\
83.62492       &   8$_{1,8}$ (v=1) $-$   7$_{1,7}$ (v=0) &  17 & 0.251 & c) \\  
83.85949       &   9$_{2,8}$ (v=0) $-$   8$_{2,7}$ (v=1) &  24 & 0.240 & d)  \\
84.43952       &   8$_{0,8}$ (v=1) $-$   7$_{0,7}$ (v=0) &  17 & 0.361 & e)  \\
85.21578$^{a}$ &   9$_{6,4}$ (v=0) $-$   8$_{6,3}$ (v=1) &  40 & 0.135 & f)  \\
85.21582$^{a}$ &   9$_{6,3}$ (v=0) $-$   8$_{6,2}$ (v=1) &  40 & 0.172 & f)  \\
85.52157       &   9$_{4,6}$ (v=0) $-$   8$_{4,5}$ (v=1) &  30 & 0.221 & g)  \\
85.60355       &   9$_{4,5}$ (v=0) $-$   8$_{4,4}$ (v=1) &  30 & 0.280 & h)  \\
86.65561       &   9$_{1,8}$ (v=0) $-$   8$_{1,7}$ (v=1) &  23 & 0.393 & i)  \\
86.66012       &   9$_{3,6}$ (v=0) $-$   8$_{3,5}$ (v=1) &  27 & 0.332 & i)  \\
87.56206       &   8$_{2,7}$ (v=1) $-$   7$_{2,6}$ (v=0) &  20 & 0.285 & j)  \\
88.36696       & 10$_{1,10}$ (v=0) $-$   9$_{1,9}$ (v=1) &  26 & 0.443 & k)  \\
88.74534       & 10$_{0,10}$ (v=0) $-$   9$_{0,9}$ (v=1) &  26 & 0.342 & l)  \\
88.81207       &   9$_{2,7}$ (v=0) $-$   8$_{2,6}$ (v=1) &  25 & 0.391 & m)  \\
88.87342$^{a}$ &   8$_{5,4}$ (v=1) $-$   7$_{5,3}$ (v=0) &  30 & 0.171 & n)  \\
88.87421$^{a}$ &   8$_{5,3}$ (v=1) $-$   7$_{5,2}$ (v=0) &  30 & 0.217 & n)  \\
88.98621       &   8$_{4,5}$ (v=1) $-$   7$_{4,4}$ (v=0) &  26 & 0.221 & o)  \\
\hline 
109.35771$^{a}$&  10$_{6,5}$ (v=1) $-$   9$_{6,4}$ (v=0) &  45 & 0.334 & p)  \\
109.35786$^{a}$&  10$_{6,4}$ (v=1) $-$   9$_{6,3}$ (v=0) &  45 & 0.422 & p)  \\
109.48919      &  10$_{5,5}$ (v=1) $-$   9$_{5,4}$ (v=0) &  40 & 0.525 & q)  \\
109.51132      &  10$_{3,8}$ (v=1) $-$   9$_{3,7}$ (v=0) &  32 & 0.536 & r)  \\
109.85770      &  10$_{4,6}$ (v=1) $-$   9$_{4,5}$ (v=0) &  35 & 0.620 & s)  \\
111.44834      &  10$_{3,7}$ (v=1) $-$   9$_{3,6}$ (v=0) &  32 & 0.716 & t)  \\
111.57944      & 11$_{1,11}$ (v=1) $-$ 10$_{1,10}$ (v=0) &  31 & 0.844 & u)  \\
112.34368      & 12$_{2,11}$ (v=0) $-$ 11$_{2,10}$ (v=1) &  40 & 0.909 & v)  \\
\hline 
168.37712      & 17$_{6,12}$ (v=0) $-$ 16$_{6,11}$ (v=1) &  93 & 1.36 & w)  \\
168.38635      & 17$_{2,15}$ (v=0) $-$ 16$_{2,14}$ (v=1) &  79 & 2.35 & w)  \\
168.44321      & 17$_{6,11}$ (v=0) $-$ 16$_{6,10}$ (v=1) &  93 & 1.69 & x)  \\
171.41795      & 16$_{6,11}$ (v=1) $-$ 15$_{6,10}$ (v=0) &  85 & 1.48 & y)  \\
\hline
\end{tabular}
\end{center}
%\vspace{-3mm}
{$^{a}$ The EG lines are autoblended, i.e., blended with other transition of EG.} \\
{$^{b}$ Integrated intensities derived from the LTE fit.} \\
\label{table-clean-transitions}
%\end{scriptsize}
\end{table*}  

%-----------------------------Table Start-----------------------------

\begin{table*}
%\begin{scriptsize}
\caption[]{Clean transitions (i.e. non blended with other molecular species) of EG identified in the interferometric SMA spectra towards G31.}
\begin{center}
%\vspace{-4mm}
\begin{tabular}{c c c c }
\hline
Frequency &  Transition &  $E_{\rm up}$  & $\int$ S$_{}\times\Delta v$   \\
(GHz) &  $J_{K_a,K_c}, v$  & (K) & (Jy/beam$\times$km s$^{-1}$) \\
\hline\hline
221.03880         &  22$_{6,17}$ (v=0) $-$ 21$_{6,16}$ (v=1)  &  143    & 8.6  \\
221.10032      & 22$_{5,18}$ (v=0) $-$ 21$_{5,17}$ (v=1) & 137 & 9.0  \\
230.57714      & 24$_{3,22}$ (v=0) $-$ 23$_{3,21}$ (v=1) & 150 & 10.0 \\
230.83032      & 24$_{2,22}$ (v=0) $-$ 23$_{2,21}$ (v=1) & 150 & 7.3 \\
231.12740      & 23$_{7,16}$ (v=0) $-$ 22$_{7,15}$ (v=1) & 160 & 7.7 \\
\hline
\end{tabular}
\end{center}
\label{table-clean-transitions-SMA}
%\end{scriptsize}
\end{table*}

\subsection{Submillimeter Array (SMA) interferometric observations}

The observations\footnote{The Submillimeter Array is a joint project between the Smithsonian Astrophysical Observatory and the Academia Sinica Institute of Astronomy and Astrophysics, and is funded by the Smithsonian Institution and the Academia Sinica (\citealt{ho04}).} were carried out in the 230 GHz band in the compact and very extended configurations on July and May 2007, respectively. The correlator was configured to a spectral resolution of 0.6 km s$^{-1}$ over both subbands, from 219.3$-$221.3 GHz (lower sideband, LSB) and 229.3$-$231.3 GHz (upper sideband, USB). The visibility data were calibrated using MIR and MIRIAD and the imaging was done with MIRIAD. The resulting synthesized beam is approximately 0.90$\arcsec$ $\times$0.75$\arcsec$ (PA=53$^{\circ}$) for the combined very extended and compact data and 1.7$\arcsec$ $\times$ 3.5$\arcsec$ (PA=59$^{\circ}$) for the compact data. For more details see \citet{cesaroni11}.
 
\begin{figure*}
\centering
\includegraphics[scale=0.5]{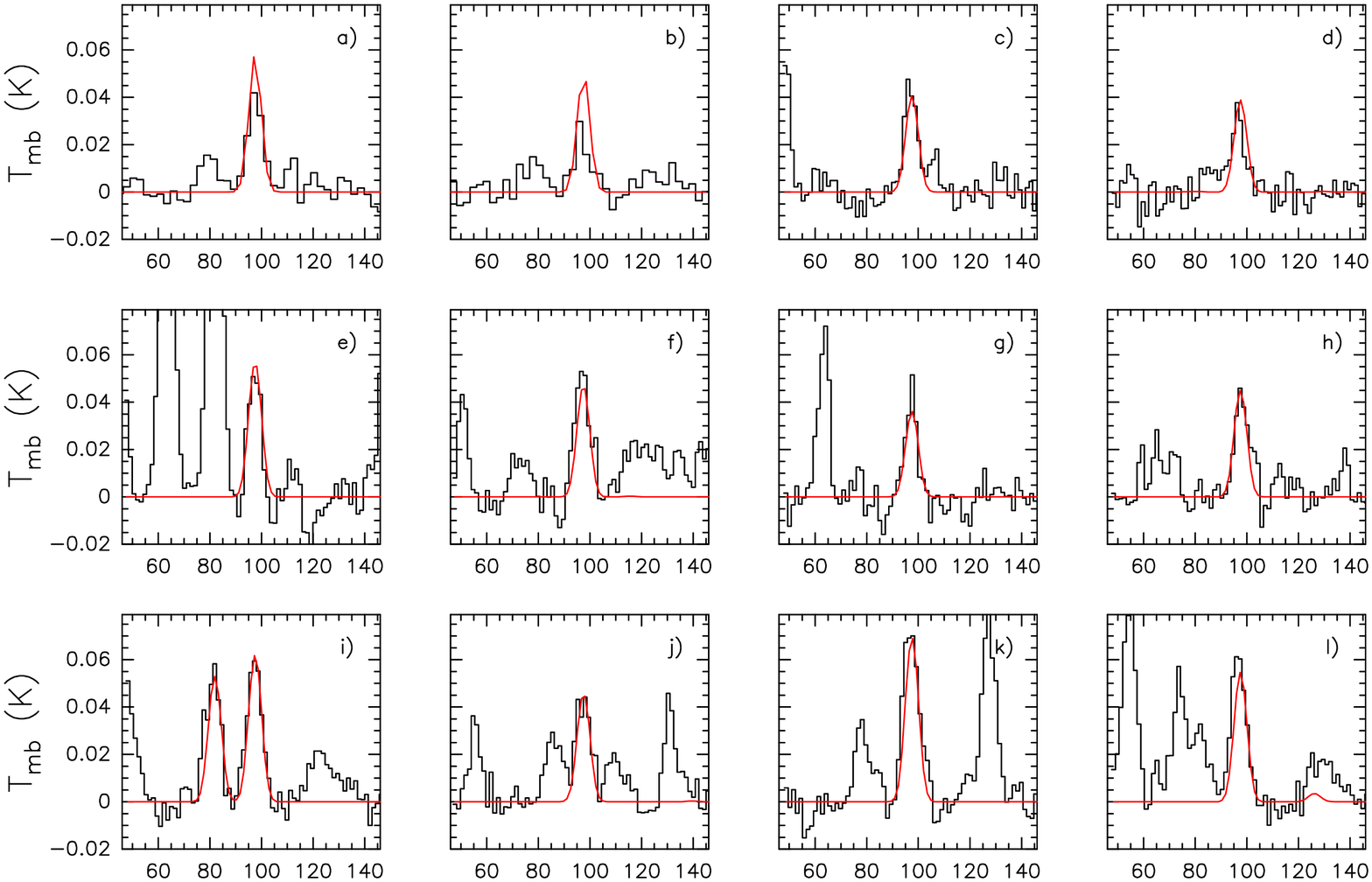}
\vskip5mm
\includegraphics[scale=0.5]{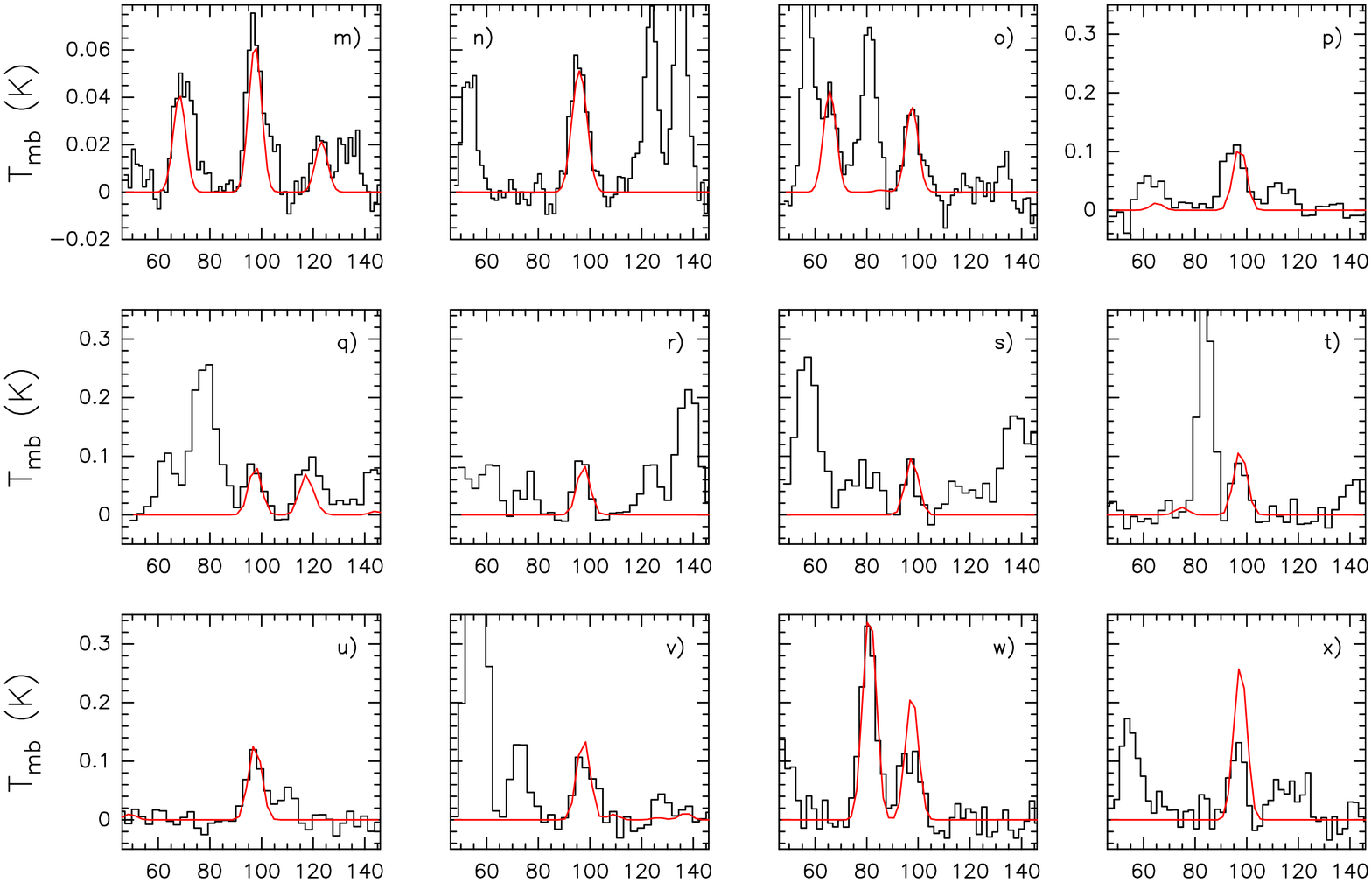}
\vskip5mm
\includegraphics[scale=0.5]{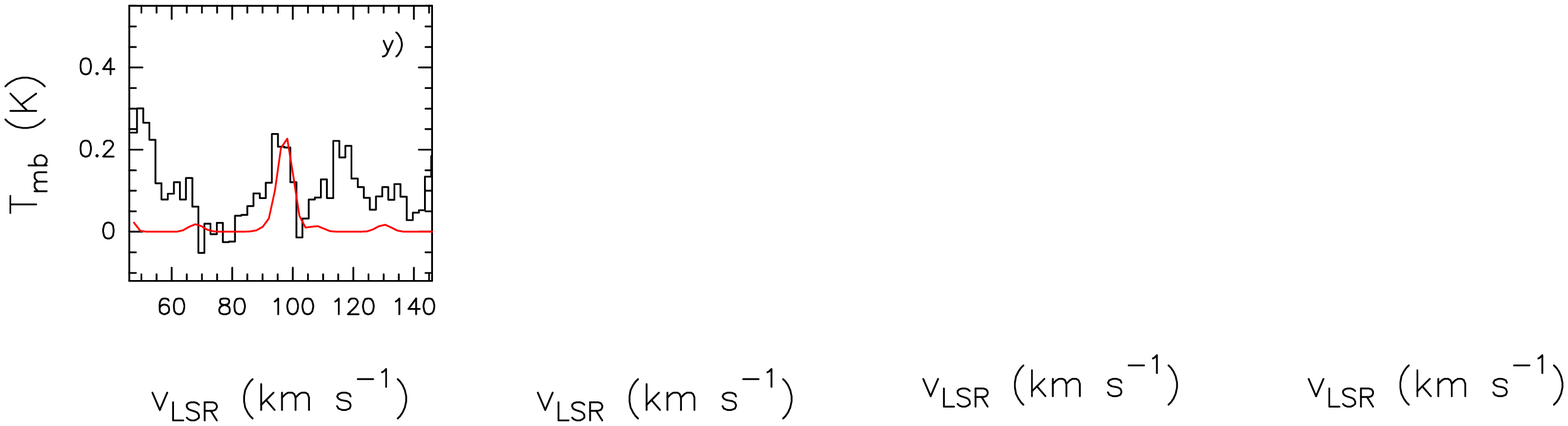}
\caption{Unblended transitions of EG from single-dish observations. The spectroscopic parameters (rest frequencies, quantum numbers, energies of the upper levels, and integrated intensities from the LTE fit are listed in Table \ref{table-clean-transitions}. For identification purposes, we have overplotted in red the LTE synthetic spectrum obtained with {\it MADCUBAIJ} assuming $T_{\rm ex}$=50 K (see text).}
\label{fig-single-clean}
\end{figure*}

\begin{figure*}
\centering
\includegraphics[scale=0.5]{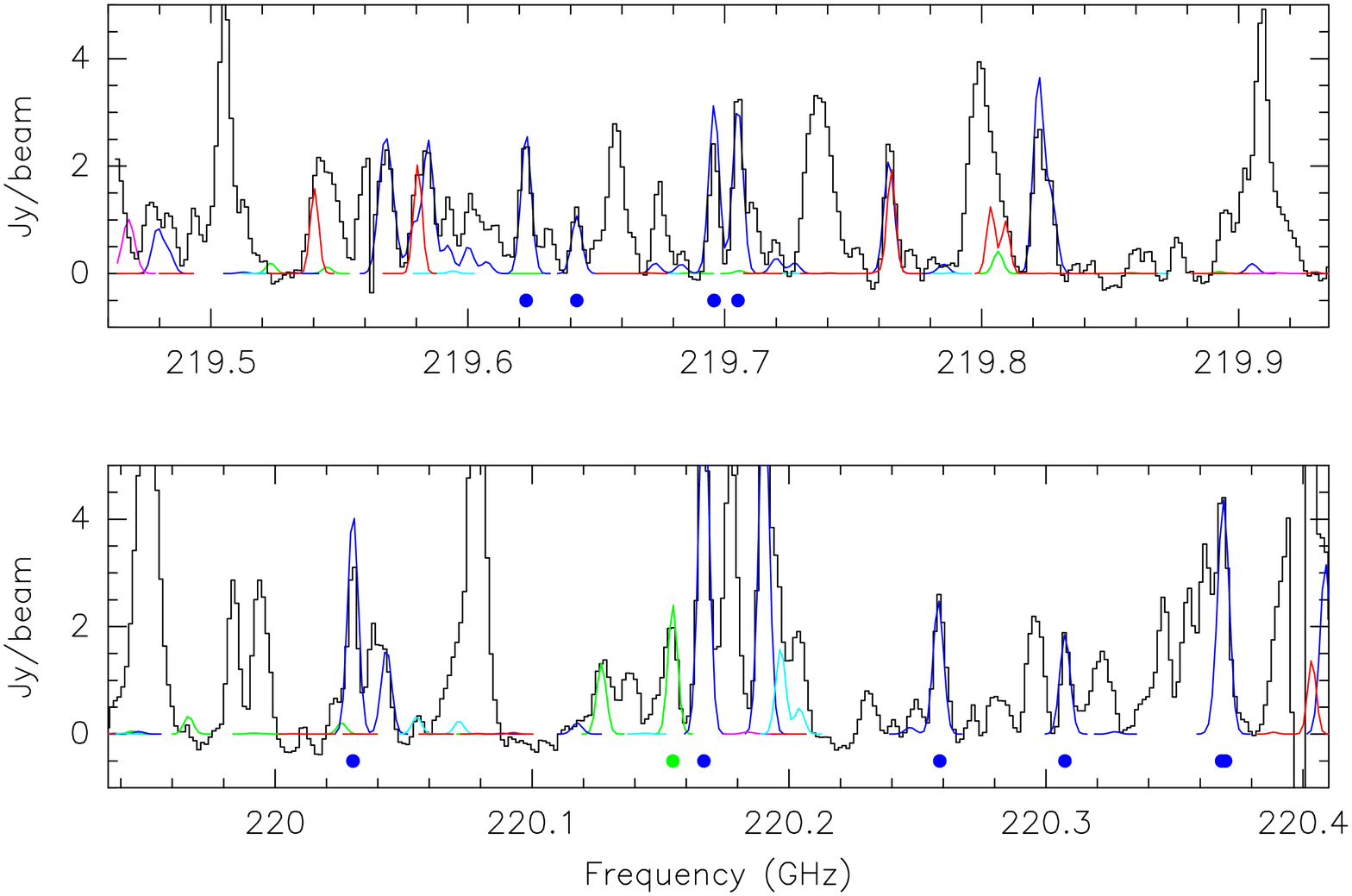}
\vskip10mm
\includegraphics[scale=0.5]{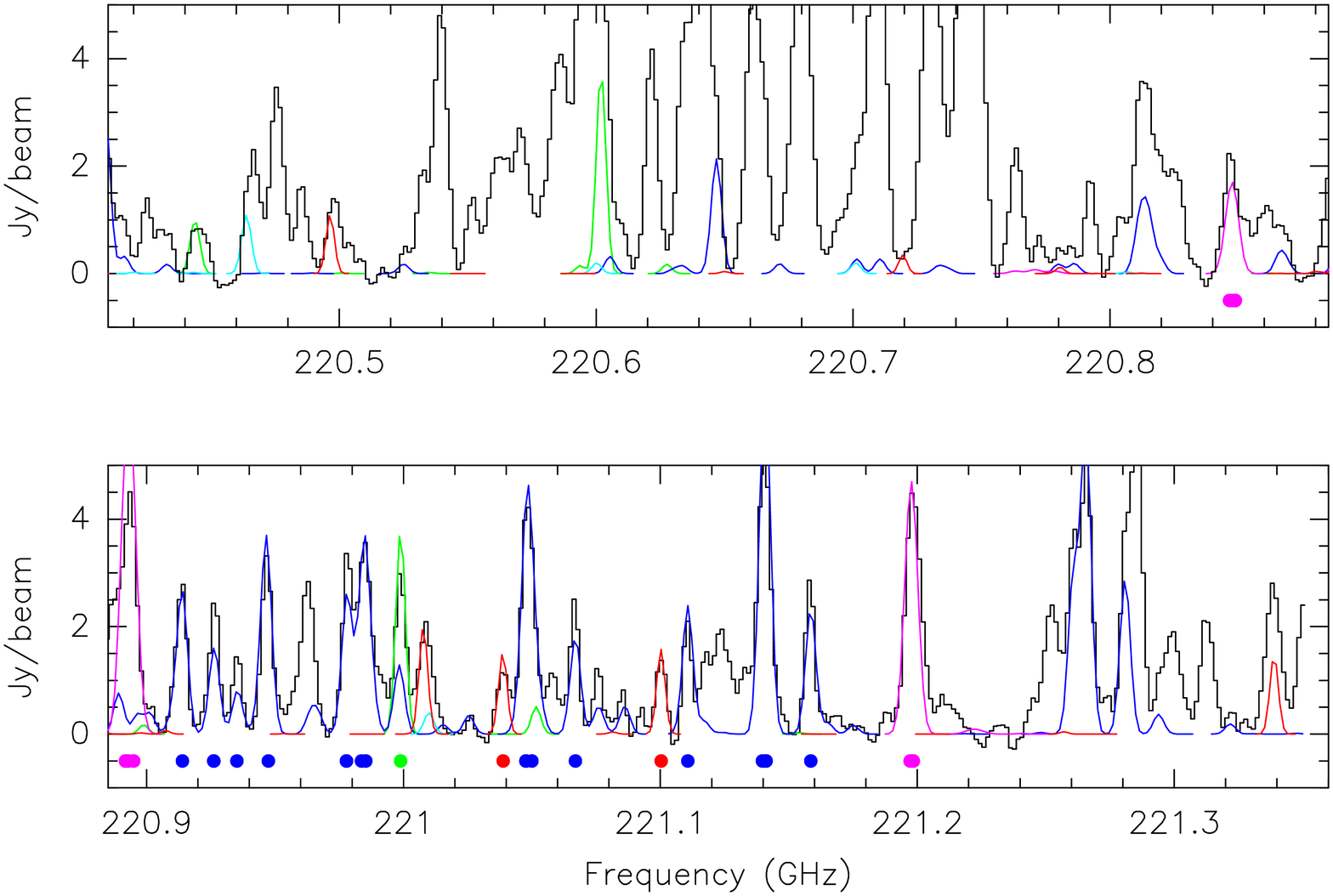}
\caption{Spectrum towards the peak of the continuum of the SMA compact data. We have overplotted the LTE synthetic spectrum of the different COMs obtained with {\it MADCUBAIJ}: EF (red), GA (light blue), MF(dark blue), DME (magenta), ET (green). The colored circles indicate the unblended lines given in Table \ref{table-clean-transitions-SMA} and Tables \ref{table-ET-SMA}, \ref{table-MF-SMA}, \ref{table-GA-SMA}, \ref{table-DME-SMA}. The physical parameters used are shown in Table \ref{table-physical-parameters-SMA}.}
\end{figure*}

\addtocounter{figure}{-1}
\begin{figure*}
\centering
\includegraphics[scale=0.5]{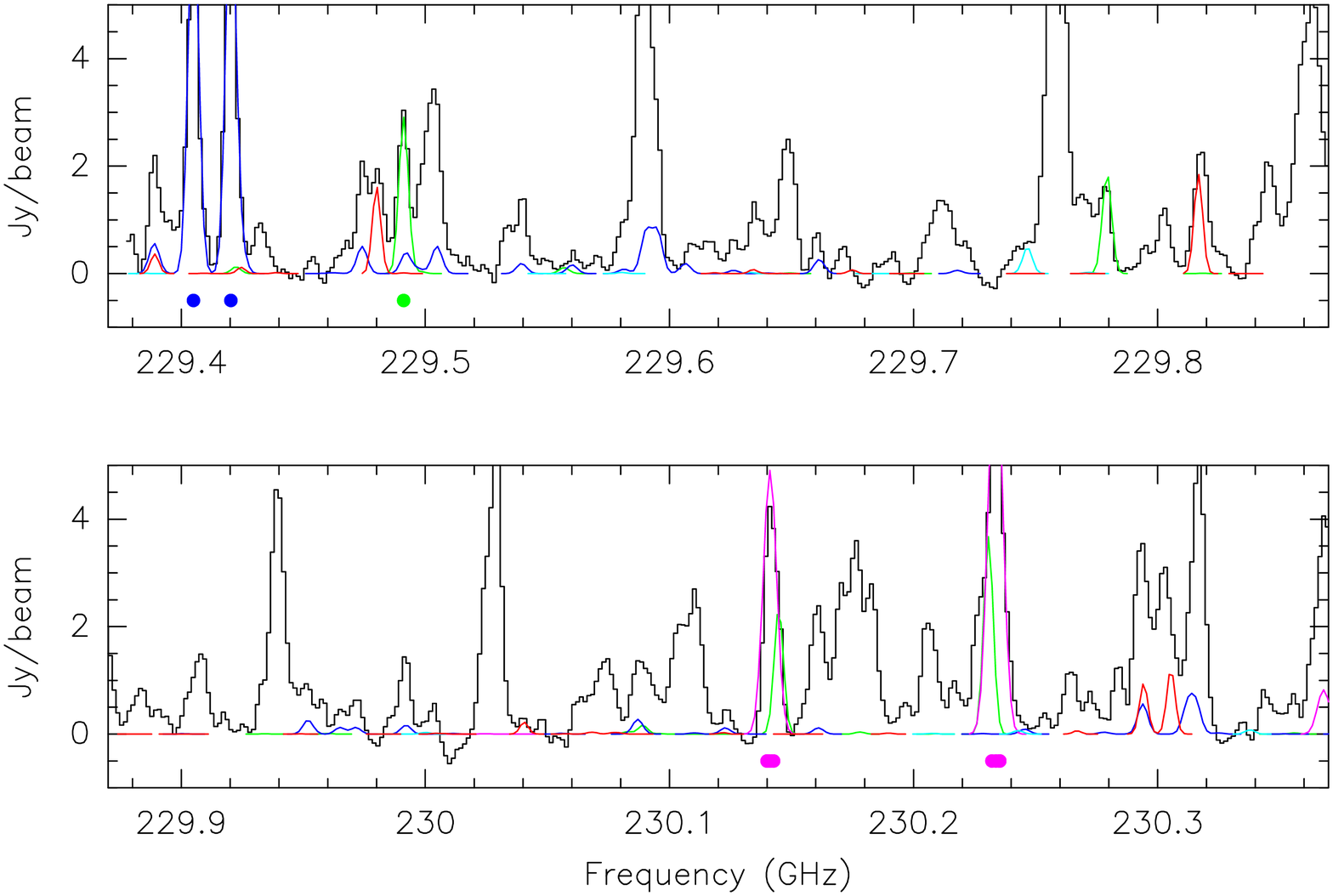}
\vskip10mm
\includegraphics[scale=0.5]{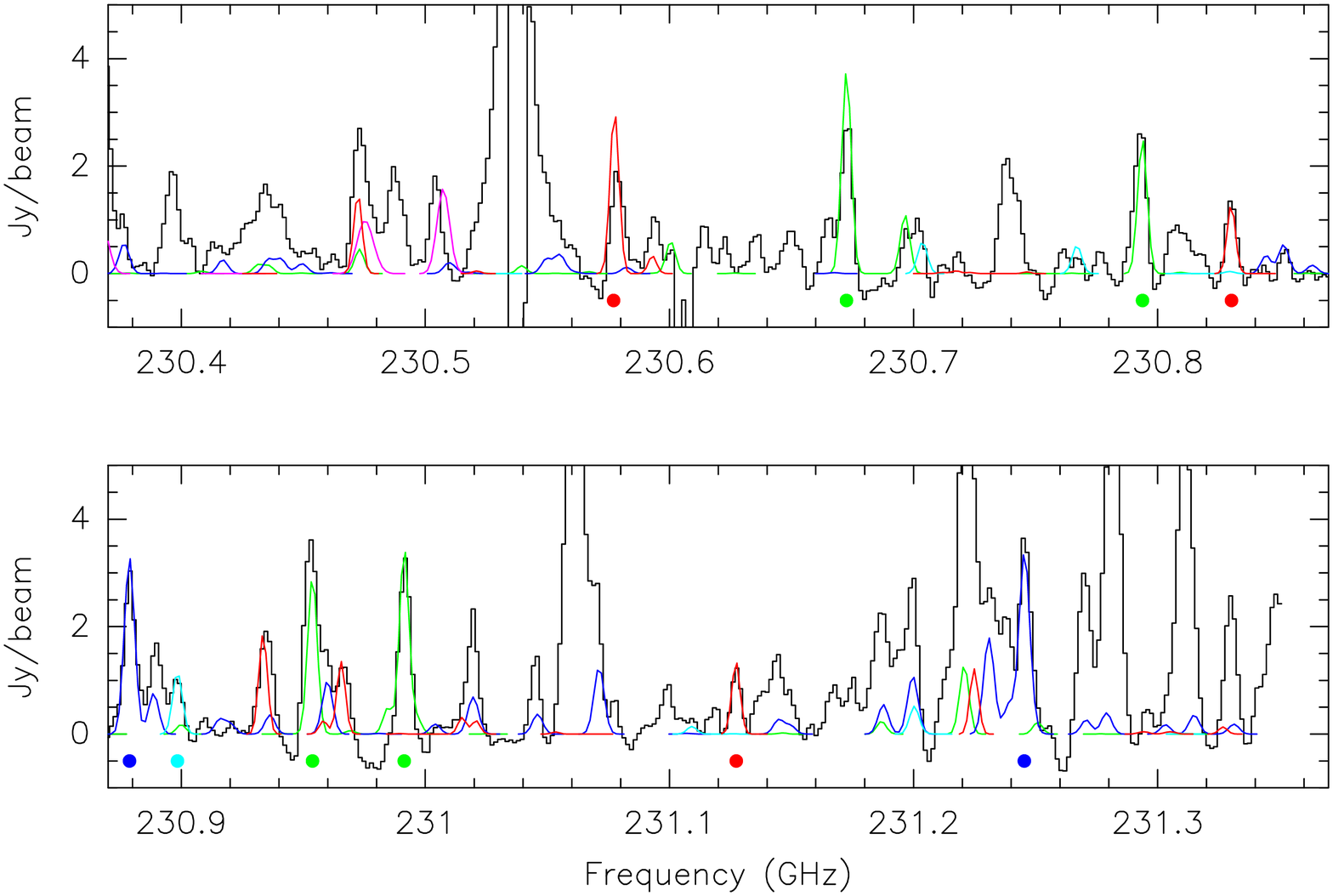}
\caption{(Continued).}
\label{figure-SMA-spectra}
\end{figure*}

\section{Line identification: ethylene glycol and other COMs}
\label{data-analysis}

A large frequency coverage is needed to robustly identify large COMs, especially in hot molecular cores where the density of lines in the spectra is high.
To assure an identification at a 99$\%$/99.8$\%$ confidence level, at least 6/8 unblended lines are needed (\citealt{halfen06}). An additional condition is that {\it all} the detectable lines predicted by a Local Thermodynamic Equilibrium (LTE) analysis that lie in the observed spectra must be revealed. 
Moreover, the relative intensities of the different transitions must be consistent with the LTE analysis, given that in the high-density environment of hot molecular cores. Furthermore, high spatial resolution data allow us to reduce the confusion limit of single-dish observations, because different molecules may exhibit different spatial distributions.

The spectra of G31 exhibits a plethora of molecular lines of multiple species. Although a complete identification and analysis of all the detected lines is planned for a forthcoming paper (Rivilla et al., {\it in prep.}), we study here the emission of selected COMs, including EG. Our motivation is to compare their physical parameters and spatial distributions. These other complex species are: glycolaldehyde (GA), ethanol (C$_2$H$_5$OH, hereafter ET), methyl formate (CH$_3$OCHO, hereafter MF) and dimethyl ether (CH$_3$OCH$_3$, hereafter DME). We will also compare the spatial distribution of these COMs with that of methyl cyanide (CH$_3$CN), a typical hot molecular tracer that has been studied in detail in \citet{cesaroni11}.

We have carried out the line identification of the COMs with {\em MADCUBAIJ}, which uses several molecular databases: JPL\footnote{http://spec.jpl.nasa.gov/} and CDMS\footnote{http://www.astro.uni-koeln.de/cdms}.
This software provides theoretical synthetic spectra of the different molecules under LTE conditions, taking into account the individual opacity of each line.

\subsection{Single-dish data: GBT and IRAM 30m}

Our wide spectral coverage has allowed us to clearly identify multiple lines of all the different COMs. In Table \ref{table-detection-summary} we summarize the molecules detected in each dataset. All the COMs have been detected in the different datasets, with the only exception of GA and DME in the GBT data, because there are no bright lines of these species in the frequency range covered by the observations.

We confirm for the first time the detection of the lowest energy conformer (aGg' conformer) of EG towards G31 (see details of the spectroscopy of EG in Appendix \ref{EG-spectroscopy}). The single-dish data (GBT+IRAM 30m) provide us with more than 30 unblended transitions of EG (see Fig. \ref{fig-single-clean} and Table \ref{table-clean-transitions}). These transitions cover an energy range of 5$-$114 K.
We note that all the transitions predicted by the synthetic spectrum above 3$\sigma$ are present in the observed spectra.

For identification purposes, we have overplotted in Fig. \ref{fig-single-clean} the simulated LTE spectra provided by {\it MADCUBAIJ}, assuming a source size from the SMA maps (see Section \ref{spatial-distribution}). The relative intensities of the lines from 7 mm to 1.3 mm are well fitted with an excitation temperature of $\sim$50 K. We note however that the LTE fit of multiwavelength transitions of EG underestimates the true temperature. The reason will be explained in detail in Section \ref{dust-absorption}. 

We have also detected multiple lines of the other COMs.
In Appendix \ref{tables-clean-transitions} we present tables with the unblended transitions detected for each species. In Figs. \ref{fig-COMs} and \ref{fig-COMs-GBT} we show selected lines at 3 mm and 7 mm detected with the IRAM 30m and GBT telescopes, respectively, along with the LTE fits obtained with {\it MADCUBAIJ} (Table \ref{table-physical-parameters-GBT}). 

\vspace{1cm}

\subsection{Interferometric data: SMA}

\begin{figure}
\centering
\includegraphics[scale=0.4]{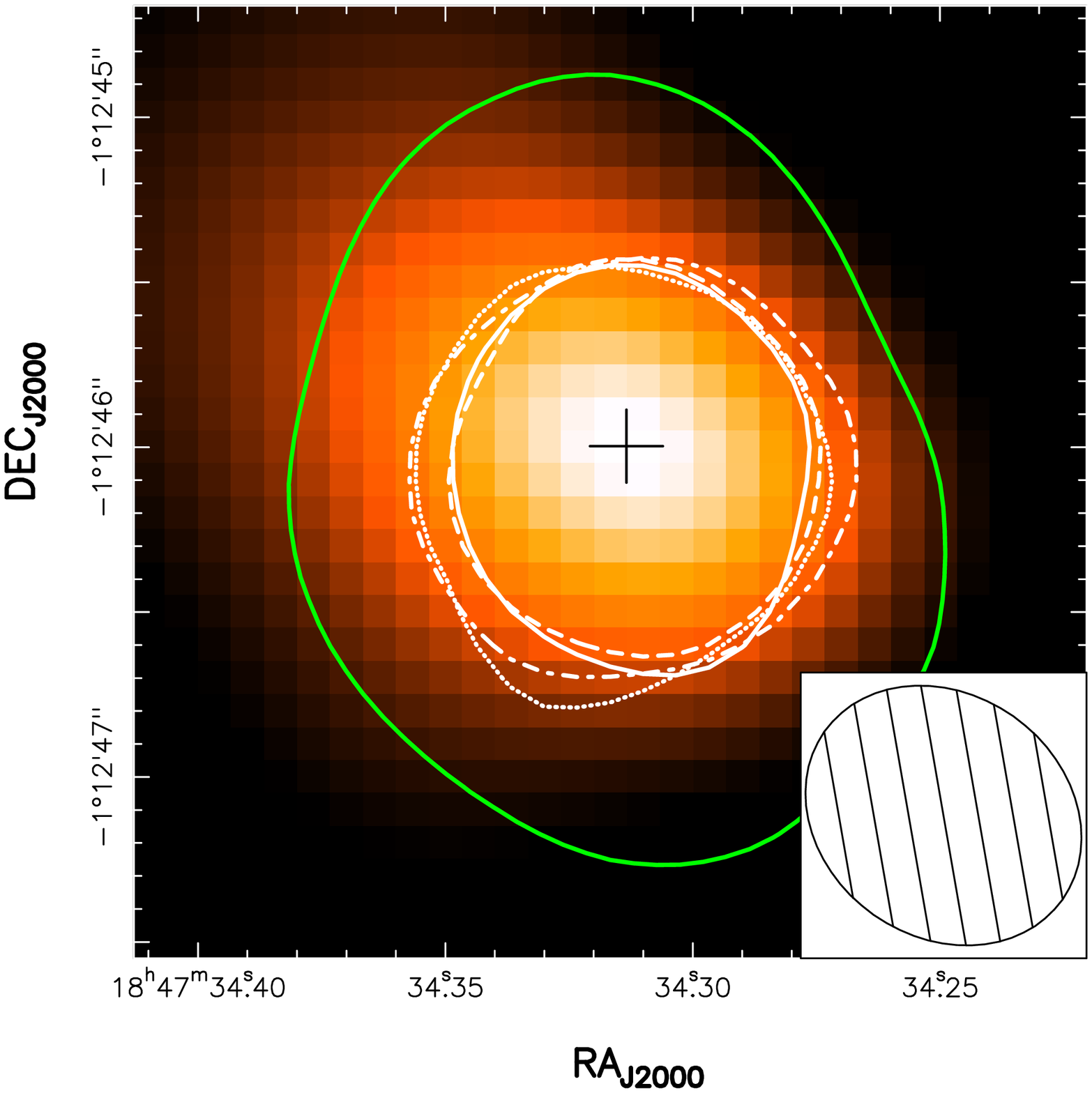}
\caption{Spatial distribution of EG transitions at 221.03880 GHz (white solid), 221.10032 GHz (white dashed), 230.57714 GHz (white dotted), and 230.830 GHZ (white dotted/dashed), overplotted on the 1.3 mm continuum observed with the SMA (color scale). The black plus sign indicates the position of the peak of the 1.3 mm continuum.  For comparison, we have also added the spatial distribution of the integrated emission of K=0,1,2 CH$_3$CN line (solid green line). All contours show the isocontour with 50$\%$ of the peak line intensity. The beam of the interferometer is shown in the lower right corner.}
\label{fig-spatial-coherence}
\end{figure}

For the identification of lines we have used the SMA datacube obtained with the compact interferometer configuration. Fig. \ref{fig-SMA-spectra} shows the full SMA spectra (containing both LSB and USB sidebands) towards the continuum peak.
For identification purposes, we have overplotted the LTE synthetic spectra obtained with MADCUBAIJ of the different COMs, and indicated with colored dots the lines least affected by blending.
Several EG unblended lines are clearly identified (Fig. \ref{fig-SMA-spectra} and Table \ref{table-clean-transitions-SMA}). The energy range covered by these transitions is 114$-$160 K.  
Additionally, the interferometric data allow us to map the spatial distribution of the EG emission. 
To study the spatial distribution of the emission we have used the datacubes obtained by combining the data of both compact and very extended configurations, which provide a better spatial resolution (0.90$\arcsec\times$0.75$\arcsec$). 
Fig. \ref{fig-spatial-coherence} confirms that the spatial distribution of different unblended EG lines is very similar. This spatial coherence supports the identification of EG.

Multiple lines of the other COMs have also been identified in the SMA data. Fig. \ref{fig-SMA-spectra} shows the LTE spectra of the different COMs. The cleanest transitions of each molecule are listed in Appendix \ref{tables-clean-transitions} (Tables \ref{table-ET-SMA}, \ref{table-MF-SMA}, \ref{table-GA-SMA}, and \ref{table-DME-SMA}).

%%%%%%%%%%%%%%%%%%%%%%%%%%%%%%%%%%%%%%%%%%%%%%%%%%%%%%%%%%%%%%%%%%%%%%%%%%%%%%%%%%%
  
\section{Derivation of physical properties}
\label{results}

%-----------------------------Table Start-----------------------------

\begin{table*}
%\begin{scriptsize}
\caption[]{Diameter of the emitting regions (see Sect. \ref{spatial-distribution}).}
\begin{center}
\vspace{2mm}
\begin{tabular}{c| c | c | c | c | c | c |  c }
\hline
 & \multicolumn{1}{c|}{EG} & \multicolumn{1}{c|}{GA} & \multicolumn{1}{c|}{ET}  & \multicolumn{1}{c|}{MF}  &   \multicolumn{1}{c|}{DME} & \multicolumn{1}{c|}{CH$_3$CN, K=0,1,2} & \multicolumn{1}{c}{CH$_3$CN, v8=1} \\ 
\hline       
$\theta_{\rm 50}$ ($\arcsec$) & 1.1 & 1.2 & 1.6 & 1.9 & 1.6 & 2.1 & 1.3  \\
$\theta_{\rm s}$ ($\arcsec$) & 0.7 & 0.9 & 1.4 & 1.7 & 1.3 & 1.9 & 1.0 \\
\hline
\end{tabular}
\end{center}
\label{table-sizes}
%\end{scriptsize}
\end{table*}   

%------------------------

\subsection{Spatial distribution of EG and other COMs} 
\label{spatial-distribution}

\begin{figure*}
\centering
\includegraphics[scale=0.6]{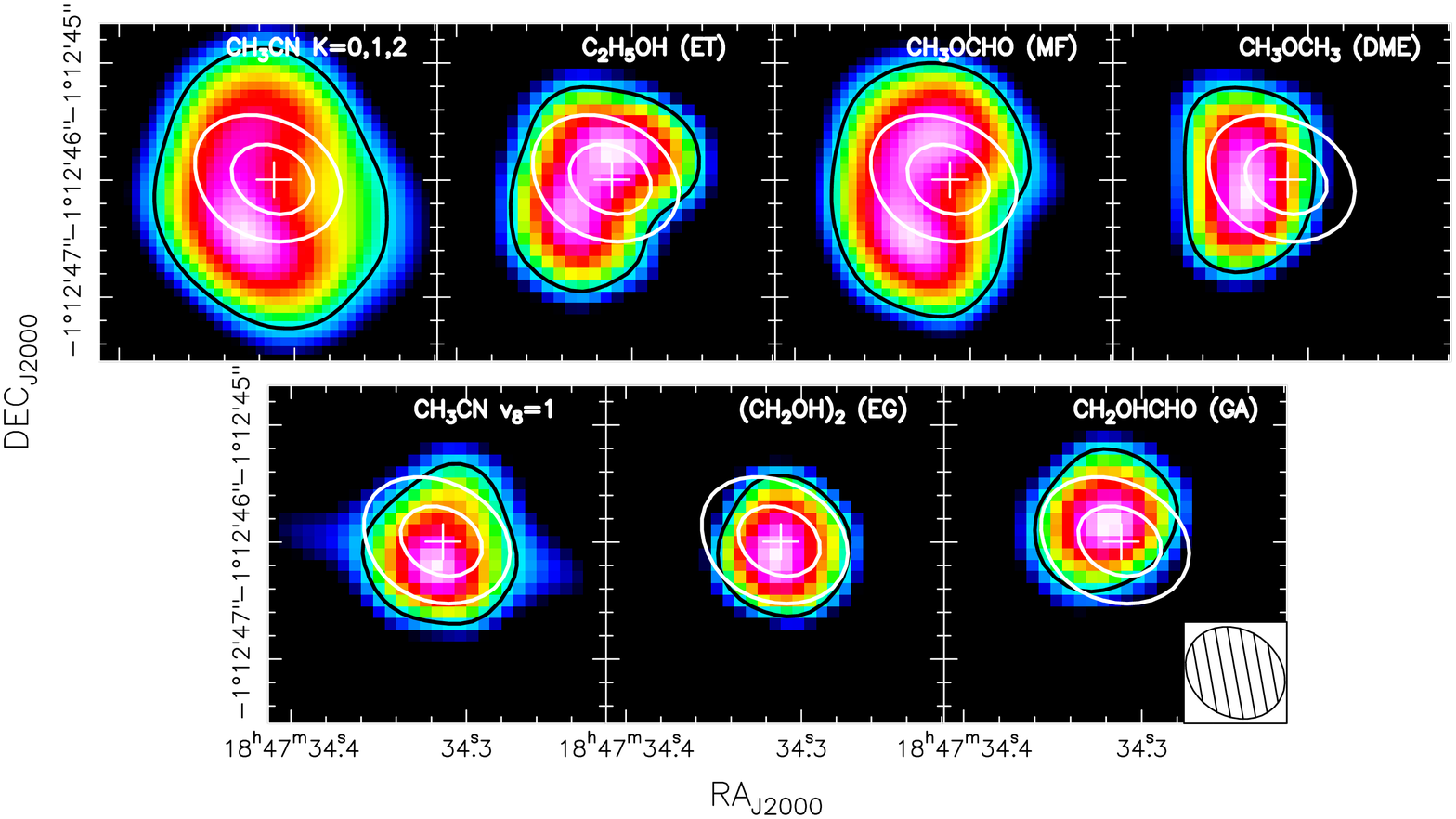}
\caption{Integrated emission maps of different COMs towards G31. All the maps have been obtained with the combination of the very extended and compact SMA data. The color scale in each panel spans from 30$\%$ to 100$\%$ of the the line intensity peak. The frequencies (and upper energies) of the lines are: 220.747 GHz (69 K), 220.743 GHz (76 K) and 220.730 GHz (97 K), for CH$_3$CN; 230.991 GHz (85 K) for ET; 229.405 GHz (111 K) and 229.420 (111 K) for MF; 230.234 (148 K) for DME; 221.228 GHz (698 K) for CH$_3$CN v$_8$=1; 221.100 (137 K) for EG; and 230.898 (131 K) for GA. The black solid line indicates the isocontour with the 50$\%$ of the peak line intensity. The two white contours correspond to the 50$\%$ and 80$\%$ of the peak value of the 1.3 mm continuum. The white plus sign indicates the positions of the peak of the 1.3 mm continuum. The beam of the interferometer is shown in the lower right corner of the lower right panel.}
\label{fig-SMA-spectra}
\label{fig-morphology}
\end{figure*}

The combined compact and very extended SMA observations have a nearly circular beam and a good uv coverage, thus allowing us to attain subarcsec resolution in all directions (0.90$\arcsec\times$0.75$\arcsec$; PA=53$^{\circ}$). Fig. \ref{fig-morphology} shows the maps of the emission of different molecules obtained by integrating the emission under unblended lines.
We have defined the deconvolved beam size of the emitting regions as $\theta_{\rm s}$=$\sqrt{\theta_{\rm 50}^{2} - \theta_{\rm beam}^{2}}$, where $\theta_{\rm beam}$ is the half-power beam width of the interferometer, and $\theta_{\rm 50}=2\sqrt{A/\pi}$ is the diameter of the circle whose area $A$ equals that enclosed inside the contour level corresponding to 50$\%$ of the line peak.

The EG emission has a diameter of 0.7$\arcsec$, and peaks towards the position of the continuum (Fig. \ref{fig-morphology}), similarly to that of GA and the vibrationally excited CH$_3$CN v$_8$=1. The EG emission is clearly smaller than the emission of other COMs like the ground vibrational level of CH$_3$CN, ET and MF (Fig. \ref{fig-morphology} and Table \ref{table-sizes}). This fact is also observed in the Orion Hot Core, where EG is more compact than MF and ET (\citealt{brouillet15}). 

The EG morphology differs from the "8-shaped" structure observed in ET, MF, or the ground state of CH$_3$CN. \citet{cesaroni11} suggested that the "8-shaped" morphology is due to line opacity, which is higher at the center of the core and hence produces the observed dip. Since more complex molecules such as EG have lower abundances, they are expected to suffer less from line opacity effects and consequently allow us to trace gas located closer to the central protostar(s). This makes EG an excellent tracer of the physical properties and kinematics of the inner regions of hot cores, which can help us to understand massive star formation itself.

%-----------------------------Table Start-----------------------------

\begin{table*}
%\begin{scriptsize}
\caption[]{Physical parameters obtained from the 3 mm IRAM 30m data for the different molecules.}
\begin{center}
%\vspace{-4mm}
\begin{tabular}{c| c| c| c| c| c }
\hline
 & \multicolumn{1}{c|}{EG} & \multicolumn{1}{c|}{ET} & \multicolumn{1}{c|}{GA}   & \multicolumn{1}{c|}{MF} &   \multicolumn{1}{c}{DME}  \\
\hline
 log$N_{\rm b}$ (cm$^{-2}$) & 15.0 & 15.9  & 14.7  & 16.2 & 15.9 \\
 log$N_{\rm s}$ (cm$^{-2}$) & 18.5 & 18.5 & 17.5  & 18.7 & 19.0 \\
\hline
$X_{\rm s}$ (10$^{-8}$)$^{a}$ & 2.6 & 0.84 & 0.26 & 4.2  & 8.4    \\
\hline
$v_{\rm LSR}$ (km s$^{-1}$) & 97.6   & 97.6 & 96.1 & 97.6 & 97.0 \\
\hline
$\Delta v$ (km s$^{-1}$) & 5.6   & 5.4 & 5.3 &  4.7 & 4.4 \\
\hline
\end{tabular}
\end{center}
\begin{scriptsize}
\vspace{-2mm}
%\hspace{5cm}
\centering
{${(a)}$ Assuming $N_{\rm H_2}$=1.2$\times$10$^{26}$ cm$^{-2}$ (see Section \ref{dust-absorption}).} \\

\vspace{-4mm}
\label{table-physical-parameters}
\end{scriptsize}
\end{table*}   

%------------------------

\subsection{The importance of dust absorption in G31}
\label{dust-absorption}

In Section \ref{data-analysis} we have fitted the 45$-$219 GHz single-dish spectra of EG using an LTE analysis with a temperature of 50 K using {\it MADCUBAIJ}. This temperature is significantly lower than that usually measured in hot cores, and in particular than that measured towards G31 by \citet{beltran05} of 164 K using CH$_3^{13}$CN and CH$_3$CN v$_{8}$=1. This discrepancy is striking, given that EG is tracing the innermost region of the core, as we have seen in the previous section, where one would expect a high temperature. Since the low excitation transitions of EG ($E_{\rm up}$=5$-$45 K) have low frequencies ($<$ 113 GHz), and the higher excitation transitions ($E_{\rm up}$=79$-$114 K) have higher frequencies ($>$168 GHz), one may wonder if such a low temperature is due to the high frequency (and energy) lines being somehow weakened with respect to the low frequency (and energy) ones. Dust absorption provide us with a frequency-dependent mechanism of this type. In fact, dust opacity increases with frequency, $\tau_\nu\sim\nu^\beta$, and hence may absorb at high frequency more effectively than at lower frequencies. This would result in an artificially low temperature when fitting with an LTE model.
 
\begin{figure}
\centering
\includegraphics[scale=0.3]{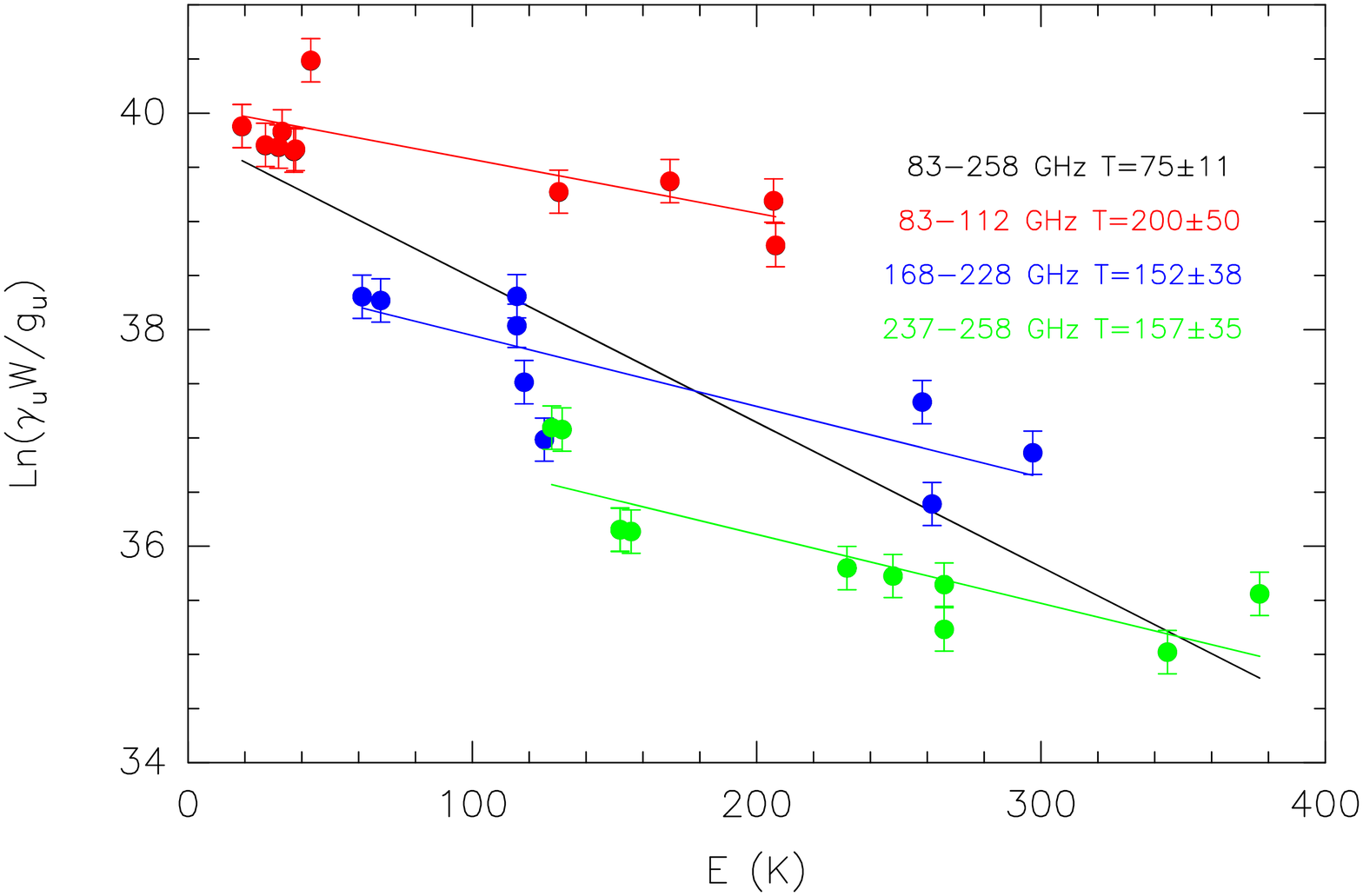}
\vskip3mm
\includegraphics[scale=0.3]{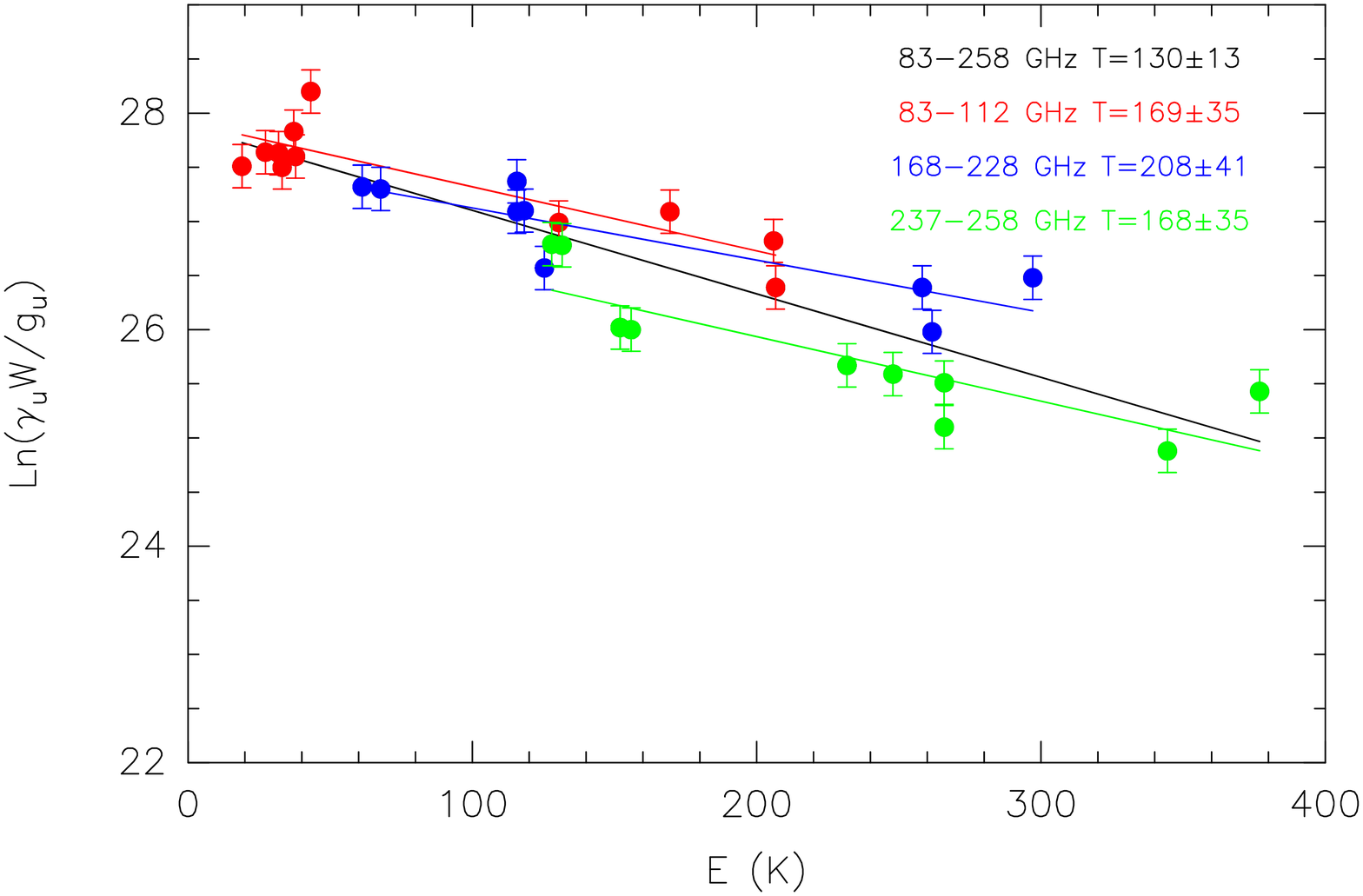}
\vskip3mm
\includegraphics[scale=0.3]{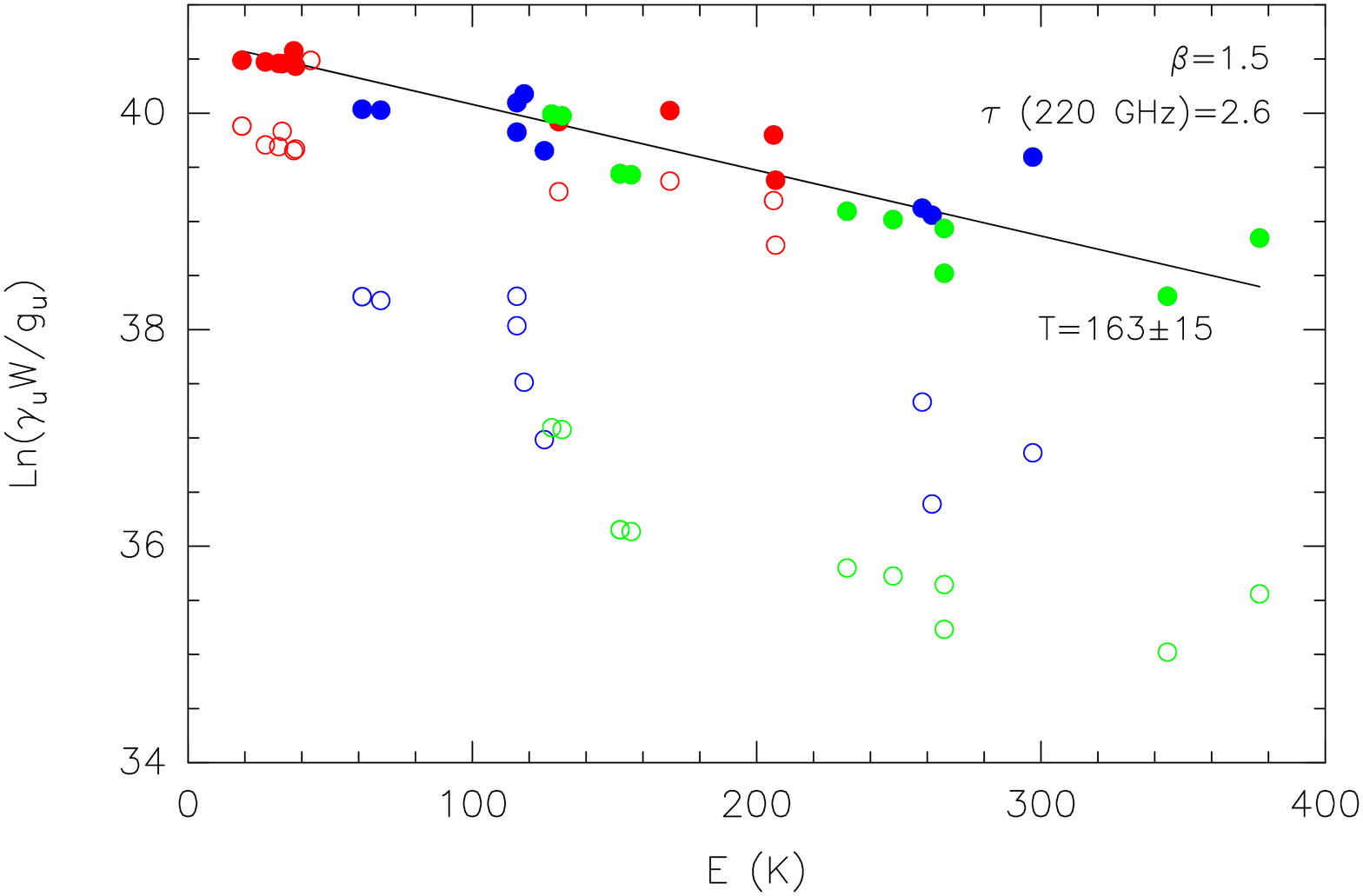}
\caption{{\it Upper panel:} Rotational diagram of MF, considering the beam dilution factor. The different colors indicate transitions in the different frequency ranges indicated with the labels. The black line is the best fit considering all transitions, while the colored lines are the best fits considering only the transitions in the same frequency range. {\it Middle panel:} Same as upper panel, without considering beam dilution, i.e., assuming that the molecular emission completely fills the telescope beam. {\it Lower panel}: Same as upper panel after correcting the line intensities for the dust opacity (solid circles) and without correction (open circles).}
\label{fig-RD}
\end{figure}

To check whether this effect is at work in the G31 core, we have studied in detail the emission of  MF, which unlike EG has the advantage of having multiple transitions covering wide ranges of $E_{\rm up}$ at similar frequencies (and hence similarly affected by dust absorption).
In the upper panel of Fig. \ref{fig-RD} we present the rotational diagram of MF (which assumes optically thin emission; see e.g. \citealt{goldsmith99}) obtained from unblended MF transitions detected with single-dish data. The values shown in the y-axis have been corrected for beam dilution, using the size derived for MF of 1.7$\arcsec$ from the SMA maps (Table \ref{table-sizes}). 
We obtained a rotational temperature of $\sim$75 K when considering all the MF transitions, similarly to the low temperature found using EG, and again lower than that expected in this hot core. 

We have repeated the analysis distinguishing in three different frequency ranges. The temperatures obtained when fitting separately transitions at similar frequencies are higher (150$-$200 K, with an average value of $\sim$170 K), in much better agreement with the value derived by \citet{beltran05} of 164 K. The upper panel of Fig. \ref{fig-RD} confirms that transitions at higher frequencies are systematically shifted to lower values in the y-axis of the rotational diagram. We interpret this result as an indication of dust opacity in G31: the line photons at high frequencies are more absorbed by dust grains than those at lower frequencies, which decreases the corresponding line intensities. 

We note that the shift in the y-axis of the rotational diagram between transitions in different frequency ranges cannot be due to different beam-dilutions at different frequencies. Even in the extreme case of extended emission filling the beam at all frequencies, the shift is still present (middle panel of  Fig. \ref{fig-RD}). However, this is not the case, because we know that both EG and MF emissions are not that extended, because our high angular resolution interferometric maps show compact sizes of 0.7$\arcsec$ and 1.7$\arcsec$, respectively. 

We suggest that the dust opacity is affecting the observed molecular line intensities. This effect is expected to be particularly important towards the innermost region of the core, where EG is detected. To support this interpretation, we have studied if the physical properties of G31 can indeed produce the dust absorption needed to explain the observed line intensities at different frequencies. 

We have derived the excitation temperature of MF considering only 3 mm transitions, which are the least affected by dust absorption. 
We have selected transitions which covers a wide range of energy levels (20$-$200 K), which allow us to constrain the excitation temperature, and with similar frequencies (in the range 83$-$88 GHz), so that the effect of dust absorption is nearly the same.
We have obtained a value of $T_{\rm ex}$=163 K, very similar to that derived by \citet{beltran05}.  

One can derive that the effect of dust opacity ($\tau_\nu$) decreases the line intensities in a factor $e^{-\rm{\tau_\nu}}$. 
Assuming a dust opacity index ($\tau_\nu\sim\nu^\beta$) typical of massive star forming cores of $\beta$=1.5 (see e.g. \citealt{palau14}), we have then calculated the value of the opacity that would provide $T_{\rm ex}$=163 K when fitting {\it all} the transitions at different frequencies, which is $\tau$(220 GHz)=2.6. The lower panel of Fig. \ref{fig-RD} shows the rotational diagram of MF after correcting the line intensities for dust opacity. 

One can derive the hydrogen column density of the core from the obtained value of the opacity using the expression:    

\begin{equation}
N_{H_2}=\frac{\tau_\nu}{\mu \hspace{1mm}  m_{\rm H} \hspace{1mm} \kappa_\nu},
\end{equation}
where $m_{\rm H}$ is the hydrogen mass, $\mu$=2.33 is the mean molecular mass per hydrogen atom, and $\kappa_\nu$ is the absorption coefficient per unit density. For $\tau$(220 GHz)=2.6, assuming $\kappa_{\rm  220 \hspace{1mm} GHz}$=0.005 cm$^{-2}$ g$^{-1}$ and a gas-to-dust ratio of 100, the derived hydrogen column density is $\sim$1.2$\times$10$^{26}$ cm$^{-2}$. Using the deconvolved angular diameter of the continuum core of $\sim$0.76$\arcsec$ (\citealt{cesaroni11}), this implies a total mass of $\sim$2300 M$_{\odot}$, which is consistent with the value of 1700 M$_{\odot}$ obtained from the flux continuum emission by \citet{cesaroni11}. Adopting the expression:

\begin{equation}
F_{\nu}=(1-e^{-\tau}) \hspace{1mm} \Omega_{\rm s} \hspace{1mm} B_{\nu}(T_{\rm d}),
\end{equation}
where $\Omega_{\rm s}$ is the beam solid angle covered by the source, and $B(T_{\rm d})$ is the Planck function. Using $\Omega_{\rm s}$=$\pi\theta_c^{2}$/4$\hspace{0.5mm}$ln2, where $\theta_c$ is the deconvolved angular diameter of the continuum core ($\sim$0.76$\arcsec$, from \citealt{cesaroni11}) and assuming $\tau$(220 GHz)=2.6 and $T_{\rm d}$=163 K, one obtains a continuum flux of 3.8 Jy, consistent with the observed value (4.6 Jy, \citealt{cesaroni11}).
Therefore, this confirms our interpretation that dust absorption is affecting the line intensities in G31. Obviously, the effect of dust absorption is expected to be important only if the level of clumpiness of the dense core is very low, i.e., if the dusty core is not fragmented into smaller condensations. Interestingly, recent high angular resolution ($\sim$0.2$\arcsec$) ALMA observations confirm that there is no hint of fragmentation in the continuum maps of G31 (Cesaroni, private communication).

In summary, a LTE analysis of multi-frequency transitions without considering the dust absorption underestimates the excitation temperatures, explaining the intriguingly low temperature of 50 K derived for EG when fitting multiwavelength transitions. In the next section we explain how to derive reliable values of the excitation temperatures and hence of molecular column densities.

%%%%%%%%%%%%%%%%%%%%%%%%%%%%%%%

\begin{figure}
\centering
\includegraphics[scale=0.375]{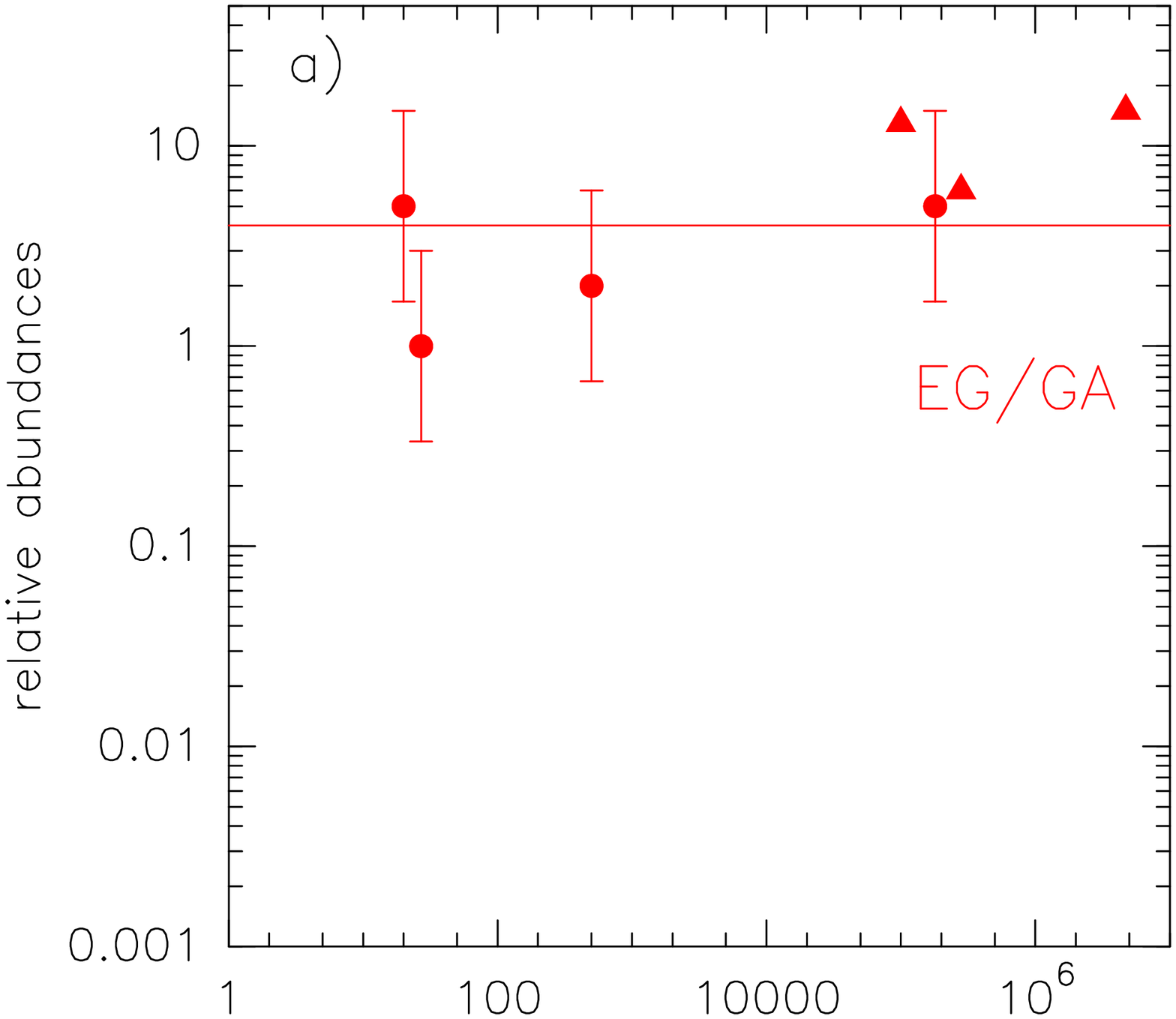}
\vskip2mm
\includegraphics[scale=0.375]{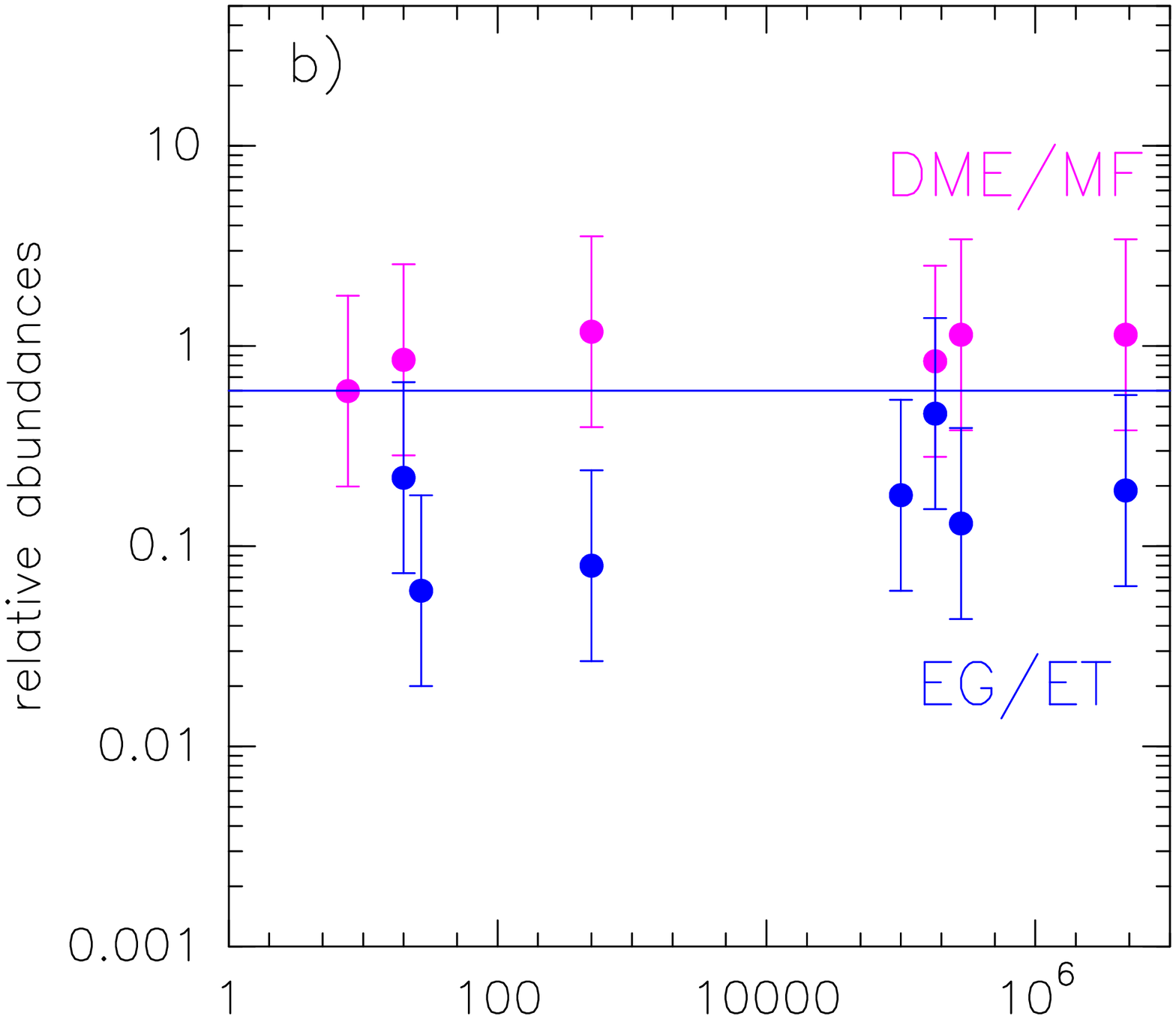}
\vskip2mm
\includegraphics[scale=0.375]{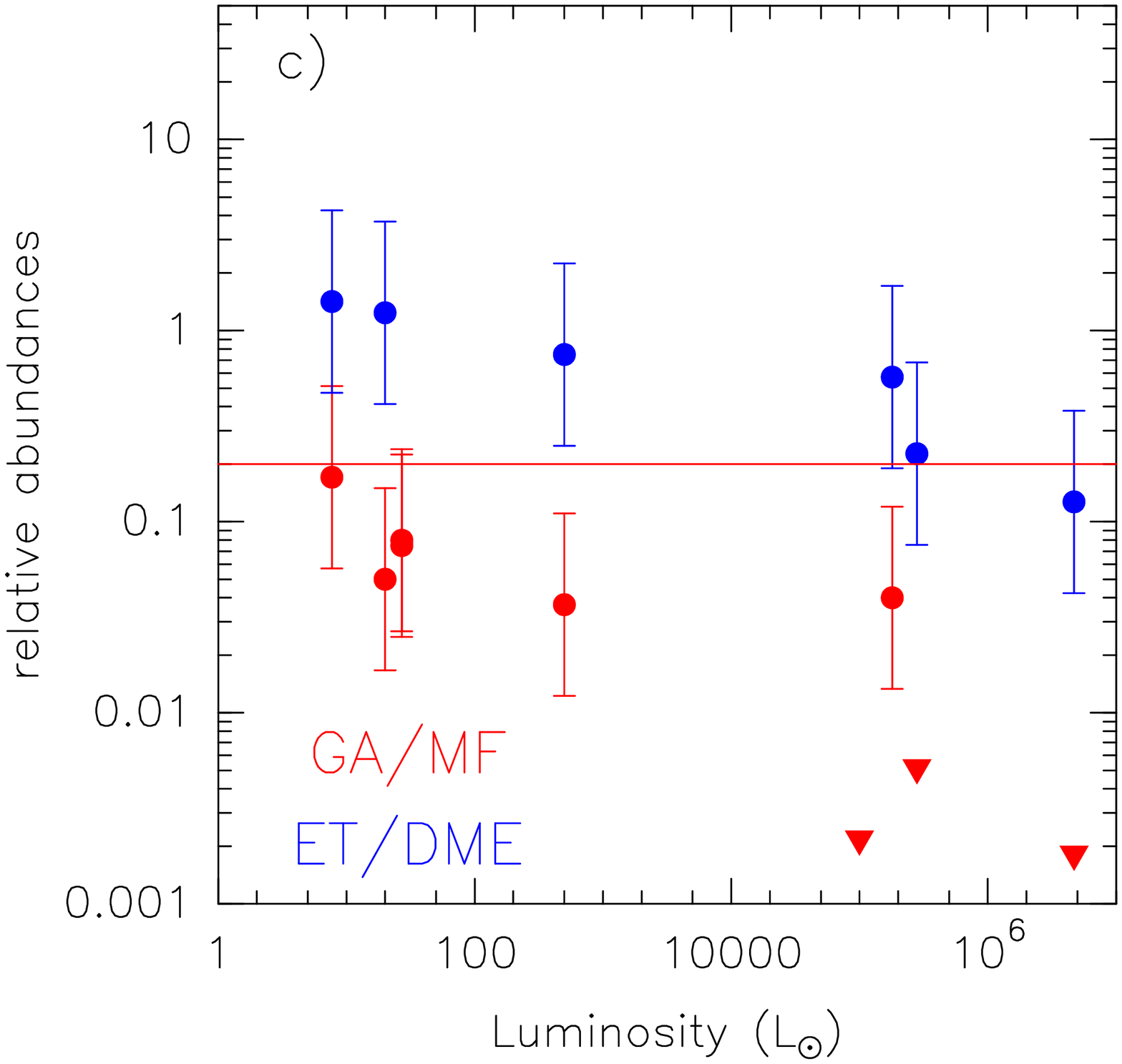}
\caption{Relative abundances of pairs of COMs with respect to the luminosity of the star-forming regions. The conservative uncertainty of the relative abundances correspond to a factor of 3 (see text). The triangles in the upper/lower panels are lower/upper limits, respectively. The solid horizontal lines correspond to the relative abundance value found in the comet Lovejoy, from \citet{biver15}.}
\label{fig-abundances}
\end{figure}

\subsection{Molecular column densities}
\label{section-physical-parameters}

With the effect of dust absorption in mind, to obtain the physical parameters of the different COMs, we have restricted our analysis only to the IRAM 30m 3 mm transitions, the least affected by the dust opacity. We have used {\it MADCUBAIJ} to fit the observed the 3 mm spectra with the LTE model\footnote{The details of the LTE synthetic spectra used to identify the different COMs in the SMA and GBT data (Figs. \ref{figure-SMA-spectra} and \ref{fig-COMs-GBT}, respectively) are presented in Appendix \ref{MADCUBA-fits}.}. 
{\it MADCUBAIJ} considers 5 different parameters to create the simulated LTE spectra: column density of the molecule ({\it N}), excitation temperature ($T_{\rm ex}$), linewidth ($\Delta$v), velocity ({\it v}) and source size ($\theta_{\rm s}$). The procedure used to derive the main physical parameters is the following: 
 
i) we have fixed the velocity of the emission and the linewidth to the values obtained from gaussian fits of unblended lines; 
ii) we have fixed the gas temperature to the value estimated using the 3 mm transition of MF (163 K), which is a good temperature tracer (\citealt{favre11});
iii) to obtain the source-average column density $N_{\rm s}$, we have fixed the size of the molecular emission to the value obtained from the integrated maps of clean lines detected with the SMA, and applied the beam dilution factor accordingly; instead we have not applied the beam dilution factor to derive the beam-averaged column density $N_{\rm b}$;
iv) leaving {\it N} as free parameter, we have used the {\it AUTOFIT} tool of {\it MADCUBAIJ} to find the value that better fits the unblended molecular transitions of each COM.

The results for the different molecular species are summarized in Table \ref{table-physical-parameters}. To account for the dust absorption discussed before, the values for the column densities have been corrected by a factor e$^{\tau(3 mm)}\sim$ 2, where  $\tau$(3 mm) has been derived assuming the derived value of $\tau$(220 GHz)=2.6 and $\beta$=1.5.

The main source of uncertainty in the estimated column densities arises from the assumption of the LTE approximation using a single temperature. However, even in the extreme case that the temperature varies by a factor of 2, the variation in the derived column densities would be less than a factor of 2. We therefore consider that the uncertainty in the estimated column densities is less than a factor of 2. 

All the molecules have velocities very close to the source systemic velocity, $\sim$97 km s$^{-1}$, and linewidths (FWHM) $\sim$5 km s$^{-1}$, typical of hot molecular cores. The derived source-averaged molecular abundances with respect to molecular hydrogen are in the range $\sim$(0.3$-$8)$\times$10$^{-8}$.

\section{Comparison with other interstellar sources}
\label{comparison}

\subsection{Comparison with other star-forming regions}

We have compared the relative molecular abundances obtained in G31 with those already reported in other star-forming regions: the massive star forming regions in Orion KL (\citealt{brouillet15}), W51e2 and G34.3+0.2 (\citealt{lykke15}); the intermediate-mass star forming region NGC 7129 FIRS2 (\citealt{fuente14}); and the low-mass protostars IRAS 16293-2422 (\citealt{jorgensen12}), NGC 1333 IRAS 2A (\citealt{maury14,coutens15,taquet15}), and NGC 133 IRAS 4A (\citealt{taquet15}). 
The different relative molecular abundances are given in Table \ref{table-abundances}. The uncertainty of the abundance ratios can be estimated by considering propagation of the uncertainty of the column densities of each molecule. We expect then that the uncertainty of relative molecular ratios is less than a factor of 3.
In Fig. \ref{fig-abundances} we show the behavior of the different abundance ratios with respect to the luminosity of the sources.

Table \ref{table-abundances} shows that EG is always more abundant than GA in all star-forming regions. This fact has also been noted by \citet{requena-torres08} in the Galactic Center, where saturated species, such as EG, are always more abundant than unsaturated species, such as GA (see Table \ref{table-abundances}); and in star-forming regions (e.g. \citealt{ikeda01}, \citealt{hollis02}, \citealt{bisschop07}) 
Therefore, there is evidence that chemistry favors saturated versus unsaturated species. \citet{requena-torres08} suggested that the double bond C$=$O (present in the unsaturated species, e.g. GA) must be easily broken.

In star-forming regions, the [EG/GA] ratio changes with luminosity, spanning from 1 to $>$15. The lower limits of the higher luminosity sources (hot cores) suggest that the [EG/GA] ratio increases with luminosity. 

The [DME/MF] and [EG/ET] ratios are nearly constant with luminosity, with values of the order of $\sim$1 and $\sim$0.1, respectively. The [GA/MF] and [ET/DME] ratios clearly decrease with increasing luminosity. In Section \ref{discussion} we will discuss how these results can help us to understand the formation of EG and other COMs.

\subsection{Comparison with cold clouds and comets}

\begin{figure}
\centering
\includegraphics[scale=0.4]{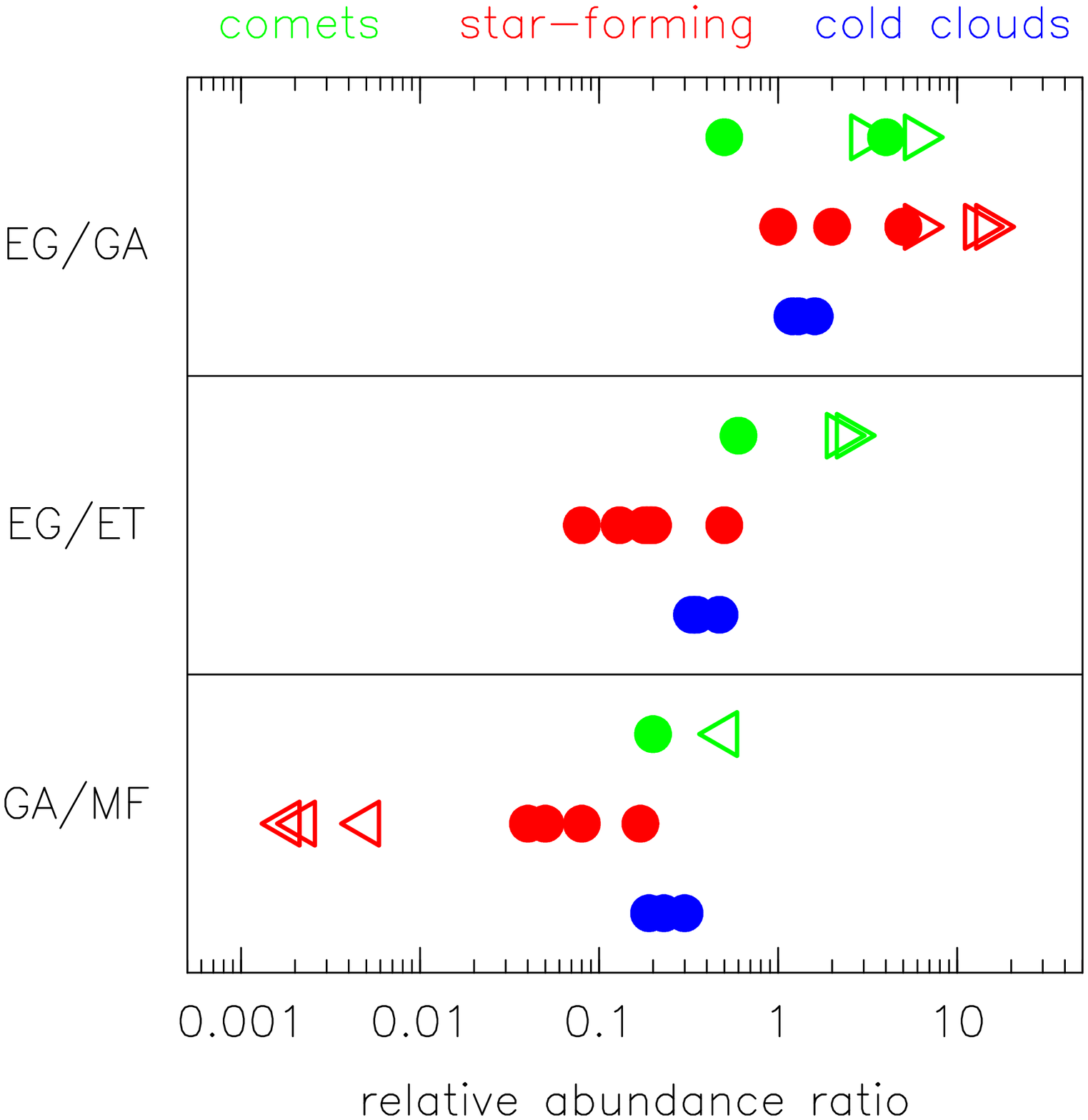}
\caption{Comparison of different ratios of molecular abundances of COMs between comets, star-forming regions and cold clouds. The color circles indicate measured ratios, while right/left-oriented triangles denote lower/upper limits.}
\label{fig-comparison-comets}
\end{figure}

The formation of stars and planets undergoes different evolutionary stages: i) cold and dense prestellar cores are able to collapse, triggering star formation; ii) the gravitational energy is converted into thermal energy and radiation, and the forming star warms up the dense neighboring envelope; iii) the envelope dissipates and a circumstellar disk remains, where eventually planets and other bodies such as comets are formed. We investigate here if there is any hint of a chemical trend between these different evolutionary stages. Namely, whether the chemical composition of star-forming regions is inherited from its proposed precursors (cold clouds), and also whether it is transferred to the bodies formed during the subsequent protoplanetary phase (e.g. comets). With this aim have also included in Table \ref{table-abundances} the relative abundance ratios towards cold clouds located in the Galactic Center, G-002, G-001, G+0.693 (\citealt{requena-torres08}), and the comets Hale-Bopp (\citealt{crovisier04a}), Lemmon (\citealt{biver14}), Lovejoy (\citealt{biver15}) and 67P/Churyumov-Gerasimenko (\citealt{goesmann15}). In Fig. \ref{fig-comparison-comets} we have compared different molecular abundance ratios between these three groups of sources (i.e. comets, star-forming regions and cold clouds).  
  
We have found that for all kind of sources [EG/GA]$>$[EG/ET]$>$[GA/MF]. This suggests that the relative abundances of different COMs, despite the observed large dispersions, share a general trend in interstellar sources at different evolutionary stages, from cold clouds to star-forming environments. This might mean that the chemistry at different stages of star formation is not significantly different, and there is a chemical thread across the entire star-formation process. However, the low number of sources with positive detections of these COMs is still low, and prevents a statistically significant conclusion.

%-----------------------------Table Start-----------------------------

\begin{table*}
%\begin{scriptsize}
\caption[]{Relative abundances of COMs in comets, star-forming regions and Galactic Center cold clouds.}
\tabcolsep 1.5pt
\begin{center}
%\vspace{-4mm}
\begin{tabular}{c| c c c c c |c | c}
\hline
 Source & [EG/GA] & [EG/ET] & [GA/MF] & [ET/DME] & [DME/MF] & Luminosity ($L_{\rm \odot}$) &  Reference \\
 \hline
\multicolumn{8}{c}{Comets} \\
\hline
Hale-Bopp & $>$6  & $>$2.5 & $<$0.5 & - & - & - & 1 \\
Lemmon  & $>$3   & $>$2.2 & $^{(a)}$ & - & -& -& 6 \\
Lovejoy  & 4  &  0.6 & 0.2 & - & - & -& 14 \\
67P  & 0.5 & - & - & - & - & - & 13 \\
\hline
\multicolumn{8}{c}{Low-mass star-forming regions (hot corinos)} \\
\hline
IRAS 16293-2422  & 1  & - & 0.08 & - & - & 2.7 & 5 \\
NGC 1333 IRAS 2A  & 5 & 0.2 & 0.05 & 1.3 & 0.9 & 20 & 7, 9, 10 \\
NGC 1333 IRAS 4A  & - & - & 0.17 & 1.4 & 0.6 & 7.7 & 10 \\
\hline
\multicolumn{8}{c}{Intermediate-mass star-forming regions} \\
\hline
NGC 7129 FIRS2 & 2 & 0.08 & 0.04 & 0.75 & 1.2 & 5$\times$10$^{2}$ & 8 \\
\hline
\multicolumn{8}{c}{High-mass star forming regions (hot cores)} \\
\hline
G31.41+0.31  & 10 & 1.0 & 0.06 & 0.3 & 2.0 & 1.8$\times$10$^{5}$ & 3, 15 \\
W51 e2 & $>$15 & 0.19 & $<$0.0018  & 0.13 & 1.1 & 4.7$\times$10$^{6}$ & 12 \\
G34.3+0.2 & $>$6 & 0.13 & $<$0.005 & 0.23  & 1.1 & 2.8$\times$10$^{5}$ & 12 \\
Orion KL & $>$13 & 0.18 & $<$0.0022 & - & - & 1$\times$10$^{5}$ & 4, 11 \\
\hline
\multicolumn{8}{c}{Galactic Center cold clouds} \\
\hline
G-002 & 1.3 & 0.33 & 0.3 & - & - & - & 2 \\
G-001 & 1.6 & 0.47 & 0.23 & - & - & - & 2 \\
G+0.693 & 1.2 & 0.35 & 0.19 & - & - & - & 2 \\
\hline
\hline
\end{tabular}
\end{center}
%\vspace{-3mm}
{$^{(a)}$ Only upper limit were reported for both species.} \\
{(1) \citet{crovisier04a}; (2) \citet{requena-torres08}; (3) \citet{beltran09}; (4) \citet{favre11}; (5) \citet{jorgensen12}; (6) \citet{biver14}; (7) \citet{maury14}; (8) \citet{fuente14}; (9) \citet{coutens15}; (10) \citet{taquet15}; (11) \citet{brouillet15}; (12) \citet{lykke15}; (13) \citet{goesmann15}; (14) \citet{biver15}; (15) this work.} \\
\label{table-abundances}
%\end{scriptsize}
\end{table*}

\section{Discussion: Formation of ethylene glycol and other COMs in star-forming regions}
\label{discussion}

We have detected 31 unblended transitions of EG with single-dish observations and 7 unblended transitions with interferometric observations. We note that the best detection up to know of EG (in SgrB2 N, the prototypical interstellar source for the search of large COMs) relies on the detection of only $\sim$13 unblended lines (\citealt{belloche13}), despite the much wider spectral coverage of these observations. This clearly shows that although the extreme conditions of the Galactic Center create a unique chemically rich environment, paradoxically the presence of so many molecules produces blending problems that makes it difficult to identify molecules with faint emission. Moreover, in SgrB2 N multiple velocity components and strong absorption also complicate a correct identification of the lines. Therefore, sources with relatively simple structure and kinematics but also chemically rich, such as G31, are a good alternative to carry out astrochemical studies. Furthermore, our work proves that single-dish observations are suitable for the detection of large COMs in hot molecular cores, despite the dilution due to the fact that the beam size is much larger than the source size. Obviously, interferometric observations at high resolution are also needed to measure the size of the emitting regions and hence obtain proper values of the molecular abundances.

The formation of COMs is still an open debate in astrophysics. Theoretical chemical models based only on gas-phase reactions have failed by orders of magnitude in reproducing the observed abundances of COMs (e.g. \citealt{geppert06}). 
Moreover, in the particular case of EG and GA there are not known gas-phase reactions to form GA or EG (see the chemical databases KIDA\footnote{http://kida.obs.u-bordeaux1.fr/} and UMIST\footnote{http://udfa.ajmarkwick.net/}; and also \citealt{woods12}).
For this reason, chemistry on the surface of dust grains has been invoked to form COMs (e.g. \citealt{garrod08}), which would be subsequently desorbed due to heating from the central protostar or from shocks, enriching the molecular environment. However, \citet{balucani15} have recently pointed out that some gas-phase reactions can be feasible, and that also gas-phase chemistry should be considered as an alternative to form COMs. 

Based on the derived molecular abundances in G31 and other star-forming regions, we discuss in this section how observations can give us clues about the formation of COMs.

We have found that the [EG/GA] ratio seems to increase with the luminosity of the source, from the low values of $\sim$1 found towards the low-mass protostar IRAS 16293-2422 (\citealt{jorgensen12} and Jorgensen et al., {\it in prep.}) to the higher value of $\sim$10 towards the hot core G31 and the lower limits of the hot cores in Orion, W51 and G34.3+0.2, which are as high as $>$ 15. 
We have also found that while the [DME/MF] and [EG/ET] are nearly constant with luminosity, the ratios between the pair of isomers [GA/MF] and [ET/DME] decrease with luminosity.
In the following subsections we propose different mechanisms that could explain these observational trends, and that give us hints about the formation of COMs in star-forming regions: i) different chemical formation routes; ii) different warm-up timescales; iii) different initial gas compositions of the dust grains; iv) different hydrogen densities of the sources; v) different formation/destruction efficiency in the gas phase.

\subsection{Chemical link}

Table \ref{table-routes} shows different chemical routes proposed in the literature to form the COMs studied in this work: EG, GA, MF, DME and ET. These chemical routes rely on theoretical chemical models (\citealt{bennett07,garrod08,woods13,balucani15}) and laboratory experiments (\citealt{fedoseev15,butscher15}). 

%-----------------------------Table Start-----------------------------

\begin{table*}
%\begin{scriptsize}
\caption[]{Chemical routes proposed in the literature for the formation of various COMs studied in this work.}
\begin{center}
%\vspace{-4mm}
\begin{tabular}{c| c c | c}
\hline
COM               &      \multicolumn{2}{c|}{Chemical pathway}   &  Reference \\
\hline
\hline
(CH$_2$OH)$_2$ (EG) & [1]    & CH$_2$OHCHO + 2H $\longrightarrow$ (CH$_2$OH)$_2$ & 4  \\
\cline{2-4}
                            & [2]    & HCO + 2H $\longrightarrow$ CH$_2$OH; &  1, 2, 6 \\
                            &        & CH$_2$OH +CH$_2$OH $\longrightarrow$ (CH$_2$OH)$_2$ &  \\

\hline
\hline
CH$_2$OHCHO (GA)    & [3]    & 2HCO $\longrightarrow$ CO + H$_2$CO $\longrightarrow$ HOCCOH; & 3 \\
                            &        & HOCCOH + H $\longrightarrow$ CH$_2$OCHO; &  \\
                            &        & CH$_2$OCHO + H $\longrightarrow$ CH$_2$OHCHO &  \\
\cline{2-4}
                            & [4]    & HCO + CH$_2$OH $\longrightarrow$ CH$_2$OHCHO & 1, 2, 6  \\
\hline
\hline
CH$_3$OCHO (MF)   & [5]    & CH$_3$O + HCO $\longrightarrow$ CH$_3$OCHO & 2 \\
\cline{2-4}
                            & [6]    & CH$_3$OCH$_2$ + O $\longrightarrow$ CH$_3$OCHO + H (gas phase) &  5 \\
\hline
\hline
CH$_3$OCH$_3$ (DME) & [7]    & CH$_3$O + CH$_3$ $\longrightarrow$ CH$_3$OCH$_3$ & 2  \\
\hline
\hline
 CH$_3$CH$_2$OH  (ET) & [8]    & CH$_2$OH + CH$_3$ $\longrightarrow$ CH$_3$CH$_2$OH & 2  \\
\hline
\end{tabular}
\end{center}
%\vspace{-3mm}
References: (1) \citet{bennett07}; (2) \citet{garrod08}; (3) \citealt{woods13}; (4)  \citet{fedoseev15}; (5) \citet{balucani15}; (6) \citet{butscher15}.
\label{table-routes}
%\end{scriptsize}
\end{table*}

If two molecular species are chemically linked, i.e., they have a common precursor and/or one is formed from the other, their relative abundance ratio should be nearly constant, regardless of the luminosity of the source. This occurs, for instance, in the [DME/MF] ratio (middle panel of Fig. \ref{fig-abundances}), which is $\sim$1 for all star-forming regions. This chemical link suggested by the observations is indeed predicted by theoretical models. \citet{garrod08} proposed that both MF and DME form from the common precursor methoxy (CH$_3$O) by surface chemistry on dust grains (route 5 and 7 in Table \ref{table-routes}, respectively). 
Complementarily, \citet{balucani15} have proposed a new gas-phase route that can efficiently contribute to the formation of MF directly from DME (route 6 in Table \ref{table-routes}). In any case, it seems that MF and DME are chemically linked, in agreement with the observed constant abundance ratio.

Likewise, if EG were formed by sequential hydrogenation of GA as suggested by the route 1, a relatively constant [EG/GA] ratio would be expected. Instead, the values of the [EG/GA] ratio differ between sources by more than an order of magnitude, from 1 to $>$15 (upper panel of Fig. \ref{fig-abundances}). This suggests that EG and GA are not directly linked.  

Therefore, it seems more plausible that EG is formed by the combination of two CH$_2$OH radicals (route 2 in Table \ref{table-routes}). Although \citet{walsh14} raised some doubts about the efficiency of this process due to the low mobility produced by the OH-bonds, the recent experimental work by \citet{butscher15} has shown that this mechanism seems to be a very efficient pathway to form EG. 

The large variation of the [EG/GA] ratio suggests not only that GA is not the direct precursor of EG, but also that they do not share a common precursor. Since we favor the formation of EG by the combination of two CH$_2$OH radicals (route 2), this would imply that CH$_2$OH is not a precursor of GA, and route 4 could be ruled out. Then, the chemical pathway based on the dimerization of HCO (route 3) proposed by \citet{woods13} appears as the most plausible scenario. This route has also been recently supported by the laboratory experiments of \citet{fedoseev15}.  

The nearly constant behavior of the [EG/ET] ratio with luminosity (middle panel of Fig. \ref{fig-abundances}) suggests that these two species might be chemically linked. Indeed, this agrees with the chemical model proposed by \citet{garrod08}, which produces EG and ET from the common precursor CH$_2$OH (routes 2 and 8, respectively).

In conclusion, the observed abundances ratios can be all justified by the routes 2 (for EG), route 3 (for GA), routes 5 and/or 6 (for MF), route 7 (for DME) and route 8 (for ET), while the routes 1 and 4 are excluded.

\subsection{Timescales}

The different warm-up timescales associated with objects of different luminosities (hot cores and hot corinos) are expected to affect the chemistry. We discuss here if this effect could explain the behavior of the relative molecular abundances with luminosity observed in Fig. \ref{fig-abundances}. According to \citet{garrod08}, the [EG/GA] and [ET/DME] ratios are higher at longer warm-up timescales, while the ratio [GA/MF] decreases with the timescale. Assuming that the warm phase in hot cores is longer than in hot corinos, as indicated by the chemical models by \citet{awad14}, then the heating timescale would be proportional to the luminosity. Hence, the timescale effect could contribute to explain the observed trends of [ET/DME] and [EG/GA] with luminosity, but not the one of [GA/MF].

\subsection{Chemical composition of dust grains}

A possible explanation for the different [EG/GA] ratios would be different chemical compositions of the dust grains in different star-forming regions. The experiments of \citet{oberg09} indicate that different initial compositions of the ices could produce very different values of the [EG/GA] ratio: pure CH$_3$OH ices produce a ratio $>$10, while ices with a composition CH$_3$OH:CO 1:10 produce a ratio $<$0.25. If this would be the explanation for the different [EG/GA] ratios, it would imply that the initial composition of grains is different for low- and high-mass star-forming regions.

\subsection{Atomic hydrogen density}

The experiments and chemical modeling by \citet{fedoseev15} show that different initial atomic H densities provide different [EG/GA] ratios. \citet{coutens15} argue that the density could explain the difference observed towards IRAS 16293-2422 and NGC 1333 IRAS 2A, with [EG/GA] ratios of 1 and 5, respectively. They suggest that the lower H$_2$ density of IRAS 2A would lead to a higher atomic H density and then to a higher [EG/GA] ratio, as observed. However, this scenario does not apply to G31, which is denser, but exhibits a [EG/GA] ratio very similar to IRAS 2A.  Therefore, we believe that atomic hydrogen density cannot explain the observed behavior of the [EG/GA] ratio.

\subsection{Formation and destruction on gas phase}

So far, we have assumed in our discussion that COMs are mainly formed by surface chemistry on dust grains and subsequently desorbed into the gas phase. However, the situation can be more complex if we also consider gas phase formation routes. Recently, \citet{balucani15} have suggested that some gas reactions may be important for the formation of MF in cold environments. If this were also the case for GA and EG, for which no efficient gas phase reactions are available to date (see \citealt{woods12}), the interpretation of the observational results should be revisited. More theoretical work is needed to make this hypothesis plausible.

Finally, another possible cause for the variation of the [EG/GA] ratio could be different destruction mechanisms in the gas phase. \citet{hudson05} show that GA is more sensitive to irradiation than EG, which may explain its lower abundance in luminous sources. Higher levels of irradiation by the higher luminosity sources would produce an increase of the [EG/GA] ratio with the luminosity, as observed. However, the destruction pathways of these COMs are still barely explored by the theoretical models and laboratory works.

\section{Summary and Conclusions}
\label{conclusions}

We have reported for the first time the detection of the aGg' conformer of the 10-atoms organic molecule ethylene glycol, (CH$_2$OH)$_2$, towards the hot molecular core G31.41+0.31. This and previous detections of COMs (e.g. CH$_2$OHCHO) confirm that G31.41+0.31 is an excellent laboratory for studying the astrochemistry related with the building blocks of primordial life. 
We have detected multiple unblended transitions of (CH$_2$OH)$_2$ with single-dish and interferometric observations. Our SMA interferometric maps show that the  (CH$_2$OH)$_2$ emission is very compact, with a diameter of 0.7$\arcsec$, peaking towards the maximum of the continuum. This morphology is similar to that observed in vibrationally excited CH$_3$CN or CH$_2$OHCHO, and differs from that observed in other more abundant COMs such as the ground state of CH$_3$CN or CH$_3$OCHO, which exhibit a dip in the central region. We propose that this difference is likely due to line opacity. Lower abundance molecules such as (CH$_2$OH)$_2$ and CH$_2$OHCHO are expectd to be more optically thin and allow us to trace the gas closest to the forming protostar(s).

Our detailed LTE analysis of (CH$_2$OH)$_2$ and CH$_3$OCHO towards G31.41+0.31 suggests that dust opacity plays an important role in the line intensities. We have found evidence that high frequency transitions are more efficiently absorbed than low frequency transitions due to higher dust opacity. This effect should be taken into account to derive a correct value of the excitation temperature and column density of molecules in dense hot molecular cores. We have derived the molecular abundances of several COMs towards G31.41+0.31 (CH$_2$OHCHO, (CH$_2$OH)$_2$, C$_2$H$_5$OH, and CH$_3$OCH$_3$), finding abundances with respect to molecular hydrogen in a range (0.3$-$8)$\times$10$^{-8}$ .

Observations in different star-forming regions indicate that the ratio (CH$_2$OH)$_{2}$/CH$_2$OHCHO spans from 1 to $>$15. We have found evidence of an increase of the ratio with the luminosity of the source. We have discussed different possible explanations for this trend, concluding that the most likely scenario is that both species are formed through different chemical routes not directly linked. We cannot exclude however a contribution of different formation and destruction efficiencies in gas phase of both species, but more laboratory works and theoretical chemical models are needed to better constrain this hypothesis.

We have discussed the molecular abundance ratios of COMs in different star-forming regions in the context of the different chemical routes proposed for their formation. We conclude that observations favor the formation of (CH$_2$OH)$_{2}$ by combination of two CH$_2$OH radicals on dust grains, in agreement with the models of \citet{bennett07} and \citet{garrod08}, and the laboratory experiments by \citet{butscher15}. The most likely formation route of CH$_{2}$OHCHO is via solid-phase dimerization of the formyl radical HCO, a mechanism proposed by \citet{woods13} and supported by the laboratory experiments of \citet{fedoseev15}. The observational data also suggests that CH$_3$OCHO and CH$_3$OCH$_3$ are chemically linked, which may mean that both share a common precursor, CH$_3$O, as proposed by \citet{garrod08}, or that the former is directly formed from the latter through viable gas-phase reactions, as suggested by \citet{balucani15}. 

The behavior of the abundance ratios of (CH$_2$OH)$_2$/CH$_2$OHCHO and C$_2$H$_5$OH/CH$_3$OCH$_3$ with luminosity may be explained by the different warm-up timescales in hot cores and hot corinos.

\begin{acknowledgements}
This work was partly supported by the Italian Ministero dell'Istruzione, Univertit\'a e Ricerca through the grant Progetti Premiali 2012 - iALMA.
\end{acknowledgements}

\bibliographystyle{mn2e}
\bibliography{biblio}

\clearpage

\appendix

\section{Spectroscopy of ethylene glycol}
\label{EG-spectroscopy}

Ethylene glycol is a triple rotor molecule. Torsions around its C$-$C bond and its two C$-$O bonds lead to 10 different conformers, four forms with an {\em anti} (A) arrangement of the OH groups and six with a {\em gauche} (G) arrangement (Fig. 1 from \citealt{christen01}). Two of the G conformers are capable of establishing intramolecular hydrogen bonds, known as g'Ga and g'Gg\footnote{The lowercase letters a and g correspond to rotation at the CO bond where clockwise/counterclockwise rotation is indicated by a positive/negative diedral angle and the symbols g/g'. The rotational directions are determined by looking along the C-C and C-O axes from the first to the second atom.}, which are expected to be energetically favorable in the gas phase.
The aGg' conformer of EG is the lowest energy state conformer. The torsion of both hydroxyl groups (OH) around the two CO bonds produces a tunneling splitting. This translates into 2 sub-levels, v=0 and v=1, separated by $\sim$7 GHz. Within each tunneling sublevel, the different rotational energy levels are characterized by three quantum numbers, $J$, $K_a$, $K_c$. To label the different energy levels we use in this paper the following notation $J_{{K_{a}},K_{c}}$, v=0 or 1. 
The transitions occur for changes from a rotational level in the v=1 state to a rotational level in the v=0 state, or viceversa. 
Figure 2 of \citet{christen95} shows a fraction of the energy level diagram of the g'Ga conformer of EG. In our analysis we have used the experimental spectrum of the aGg' conformer obtained by \citet{christen95} and \citet{christen03}. 

\section{Identified unblended transitions of COMs}
\label{tables-clean-transitions}

We present in this appendix several tables with the unblended transitions of the different COMs found in the single dish and interferometric data: ET (Tables \ref{table-ET-single-dish} and \ref{table-ET-SMA}), MF (Tables \ref{table-MF-single-dish} and \ref{table-MF-SMA}), GA (Tables \ref{table-GA-single-dish} and \ref{table-GA-SMA}) and DME (Tables \ref{table-DME-single-dish} and \ref{table-DME-SMA}).

\begin{figure*}
\centering
\includegraphics[scale=0.45]{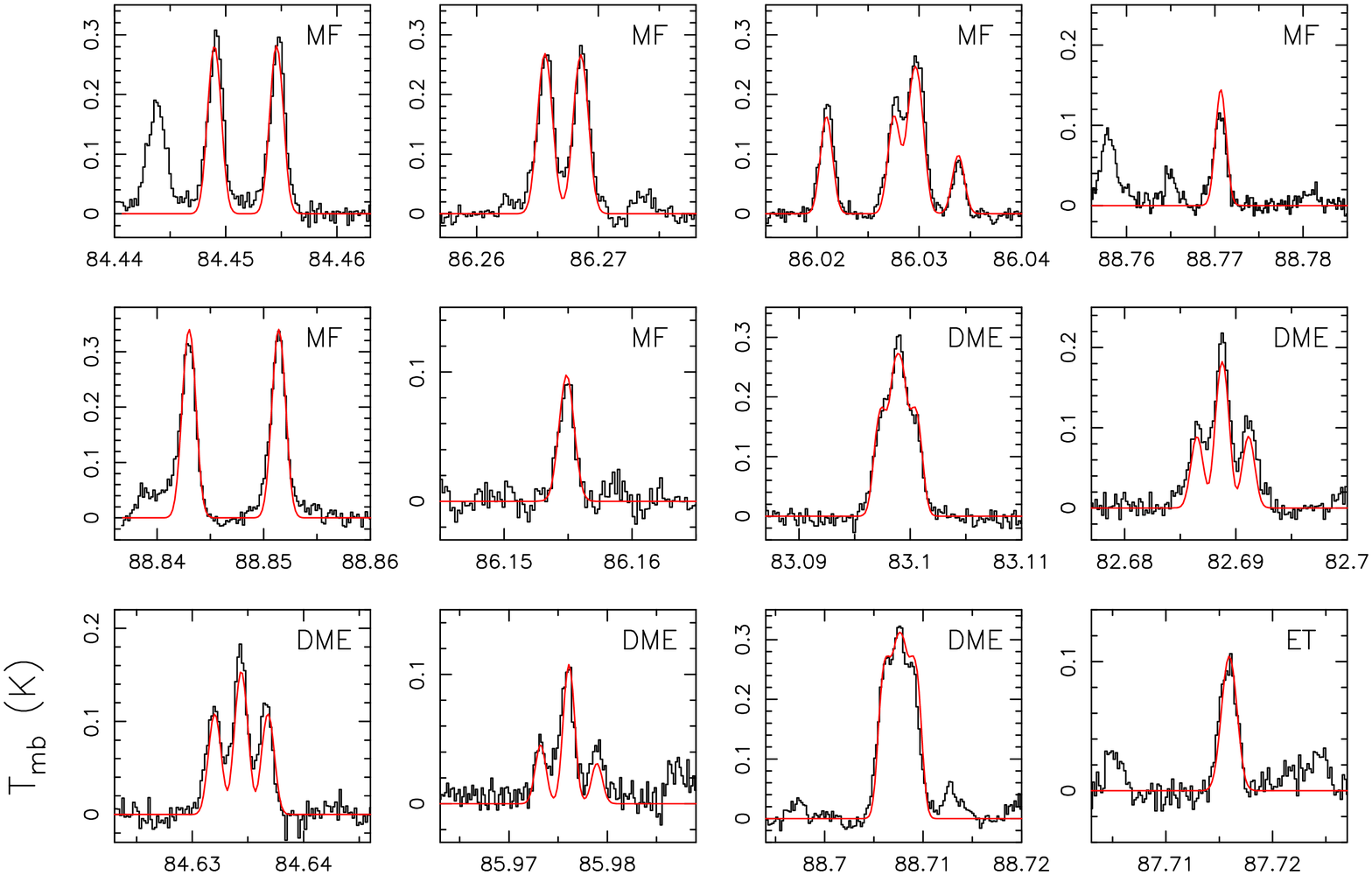}
\vskip4mm
\hspace{5mm}
\includegraphics[scale=0.45]{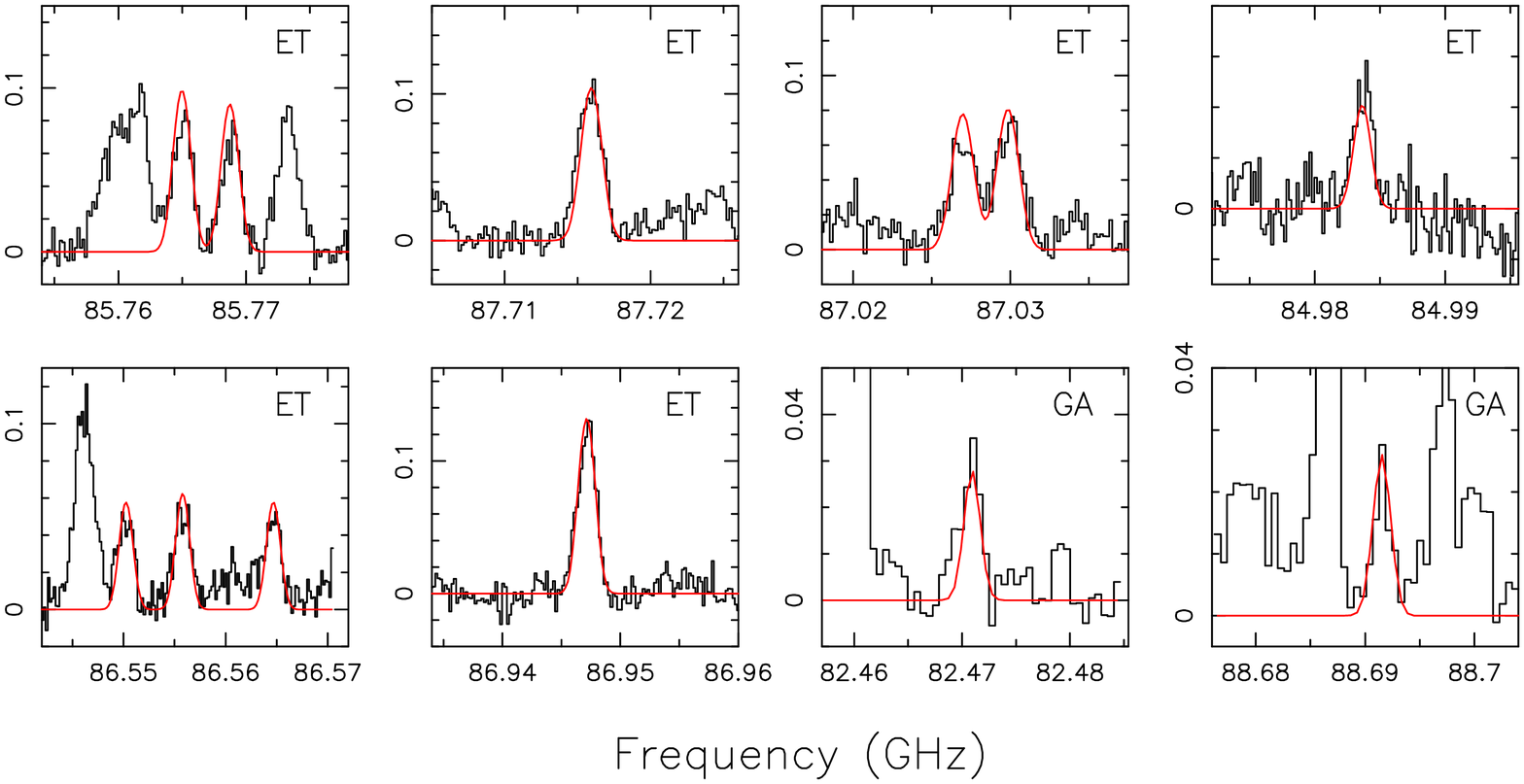}
\caption{Selected transitions at 3 mm of COMs detected towards the G31.41+0.31 hot molecular core with the IRAM 30m telescope: methyl formate (MF), dimethyl ether (DME), ethanol (ET) and glycolaldehyde (GA). We have overplotted in red the LTE synthetic spectra obtained with {\it MADCUBAIJ} using the physical parameters given in Table \ref{table-physical-parameters}.}
\label{fig-COMs}
\end{figure*}

In Fig. \ref{fig-COMs} we also present some selected IRAM 30m spectra at 3 mm of MF, ET, DME and GA. We have overplotted in red the LTE fit using {\it MADCUBAIJ} with the physical parameters shown in Table \ref{table-physical-parameters}.

\begin{figure*}
\centering
\includegraphics[scale=0.45]{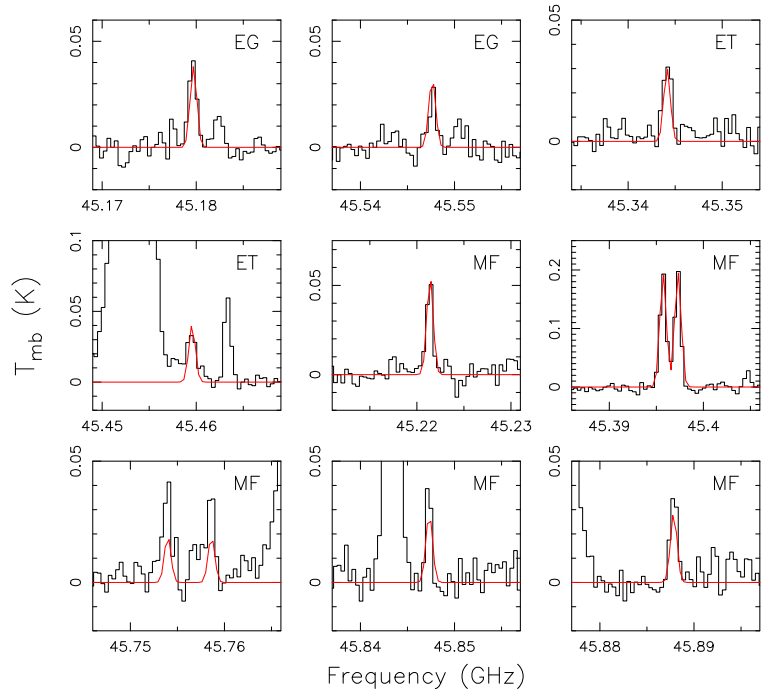}
\caption{Unblended transitions of the COMs detected towards the G31.41+0.31 hot molecular core with the GBT telescope: ethylene glycol (EG), ethanol (ET) and methyl formate (MF). We have overplotted in red the LTE synthetic spectra obtained with {\it MADCUBAIJ} using the physical parameters given in Table \ref{table-physical-parameters-GBT}.}
\label{fig-COMs-GBT}
\end{figure*}

In Fig. \ref{fig-COMs-GBT} we present the unblended transitions of the COMs detected with the GBT telescope.

%-----------------------------Table Start-----------------------------

\begin{table*}
%\begin{scriptsize}
\caption[]{Unblended transitions of ET identified in the single dish spectra (GBT and IRAM 30m) towards G31.}
\begin{center}
%\vspace{-4mm}
\begin{tabular}{c c c }
\hline
Frequency &  Transition$^{a}$ &  $E_{\rm up}$  \\
(GHz) &   & (K)  \\
\hline\hline
45.34445     & 4(2,2)$-$3(1,2), v$_{\rm t}$=0$-$1   & 70  \\
45.45987     & 13(3,11)$-$12(4,8)  &  88 \\
81.68343    & 8(1,7)$-$7(2,6)  & 32   \\
83.56877    & 5(1,5)$-$4(1,4), v$_{\rm t}$=0$-$0 & 70   \\
83.66196    & 5(1,5)$-$4(1,4), v$_{\rm t}$=1$-$1  & 75  \\
84.59587    & 4(2,3)$-$4(1,4)  & 13   \\
85.26550    & 6(0,6)$-$5(1,5)  &  17 \\
85.74718    & 5(0,5)$-$4(0,4), v$_{\rm t}$=0$-$0  & 69   \\
85.76515    & 8(0,8)$-$8(1,8), v$_{\rm t}$=1$-$0  & 91   \\
85.76890    & 5(0,5)$-$4(0,4), v$_{\rm t}$=1$-$1  & 74   \\
86.31130    & 5(2,4)$-$4(2,3), v$_{\rm t}$=0$-$0  & 74   \\
86.51654    & 5(4,2)$-$4(4,1), v$_{\rm t}$=1$-$1  & 94  \\
86.51654    & 5(4,1)$-$4(4,0), v$_{\rm t}$=1$-$1  & 94  \\
86.55042    & 5(3,3)$-$4(3,2), v$_{\rm t}$=1$-$1  & 85   \\
86.55596    & 5(4,2)$-$4(4,1), v$_{\rm t}$=0$-$0  & 89 \\
86.55596    & 5(4,1)$-$4(4,0), v$_{\rm t}$=0$-$0  & 89   \\
86.56484    & 5(3,2)$-$4(3,1), v$_{\rm t}$=1$-$1  & 85   \\
86.62171    & 5(3,2)$-$4(3,1), v$_{\rm t}$=0$-$0  & 80 \\
87.02713    & 5(2,3)$-$4(2,2), v$_{\rm t}$=1$-$1  & 79  \\
87.03001    & 5(2,3)$-$4(2,2), v$_{\rm t}$=0$-$0  & 74  \\
87.71610    & 5(2,4)$-$5(1,5)  & 18  \\
88.69737    & 8(1,7)$-$7(0,7), v$_{\rm t}$=0$-$1  & 89   \\
98.58509    & 15(1,15)$-$15(0,15), v$_{\rm t}$=1$-$0  & 158 \\
112.12954    & 12(3,9)$-$12(2,10)   & 77   \\
140.50681    & 8(2,6)$-$7(2,5), v$_{\rm t}$=0$-$0  & 92   \\
147.42746    & 8(1,8)$-$7(0,7)  & 30   \\
154.93019    & 9(2,8)$-$8(2,7), v$_{\rm t}$=1$-$1  & 104   \\
168.12339    & 10(0,10)$-$9(0,9), v$_{\rm t}$=0$-$0  & 102  \\
168.24791    & 10(0,10)$-$9(0,9), v$_{\rm t}$=1$-$1  & 106   \\
169.70219    & 18(3,16)$-$18(2,17)  & 155   \\
171.96575    & 10(2,9)$-$9(2,8), v$_{\rm t}$=1$-$1  & 112   \\
173.07117    & 10(6,4)$-$9(6,3), v$_{\rm t}$=1$-$1  & 152  \\
173.07117    & 10(6,5)$-$9(6,4), v$_{\rm t}$=1$-$1  & 152   \\
173.24010    & 10(5,6)$-$9(5,5), v$_{\rm t}$=0$-$0  & 133   \\
173.24010    & 10(5,5)$-$9(5,4), v$_{\rm t}$=0$-$0  & 133   \\
173.39133    & 5(2,3)$-$4(1,4)  & 18   \\
173.39339    & 10(4,6)$-$9(4,5), v$_{\rm t}$=0$-$0  & 122   \\
174.23291    & 6(2,5)$-$5(1,4)  & 23   \\
224.44306    & 12(7,6)$-$12(6,6), v$_{\rm t}$=0$-$1  & 181  \\
224,44311    & 12(7,5)$-$12(6,7), v$_{\rm t}$=0$-$1  & 181   \\
224.59698    & 16(7,10)$-$16(6,10), v$_{\rm t}$=0$-$1  & 230  \\
224.59850    & 16(7,9)$-$16(6,11), v$_{\rm t}$=0$-$1  & 230   \\
224.66995    & 3(2,1)$-$2(1,1), v$_{\rm t}$=1$-$0  &  71  \\
224.73929    & 11(3,9)$-$10(2,9), v$_{\rm t}$=0$-$1  & 123   \\
224.82313    & 13(1,12)$-$12(1,11), v$_{\rm t}$=0$-$0  & 135  \\
228.02905    & 13(3,10)$-$13(3,9), v$_{\rm t}$=0$-$0  & 144   \\
\hline
\end{tabular}
\end{center}
\label{table-ET-single-dish}
%\end{scriptsize}
\end{table*}

  %-----------------------------Table Start-----------------------------

\begin{table*}
%\begin{scriptsize}
\caption[]{Unblended transitions of ET identified in the interferometric SMA spectra towards G31.}
\begin{center}
%\vspace{-4mm}
\begin{tabular}{c c c  }
\hline
Frequency &  Transition$^{a}$ &  $E_{\rm up}$   \\
(GHz) &  & (K)  \\
\hline\hline
220.15478     & 24(3,22)$-$24(2,23) & 263 \\
220.99889     & 13(0,13)$-$12(0,12), v$_{\rm t}$=1$-$1 & 135.5   \\
229.49113     & 17(5,12)$-$17(4,13) & 160.1   \\
230.67255     & 13(2,11)$-$12(2,10), v$_{\rm t}$=0$-$0 & 138.6  \\
230.79386     & 6(5,2)$-$5(4,2), v$_{\rm t}$=0$-$1 & 104.8   \\
230.95378     & 16(5,11)$-$16(4,12) & 145.8   \\
230.99137     & 14(0,14)$-$13(1,13)  & 85.5   \\
\hline
\end{tabular}
\end{center}
\vspace{2mm}
$^{a}$ {The transitions of the {\it gauche} state of ethanol are designated with $v_{\rm t}$=0 ({\it gauche +}) and $v_{\rm t}$=1 ({\it gauche -}), while the transions without $v$ number correspond to the {\it trans} state.}
\label{table-ET-SMA}
%\end{scriptsize}
\end{table*}

%%%%%%%%% MF
%-----------------------------Table Start-----------------------------

\begin{table*}
%\begin{scriptsize}
\caption[]{Unblended transitions (i.e. non blended with other molecular species) of MF identified in the single dish spectra (GBT and IRAM 30m) towards G31.}
\begin{center}
%\vspace{-4mm}
\begin{tabular}{c c c }
\hline
Frequency &  Transition &  $E_{\rm up}$  \\
(GHz) &    & (K)  \\
\hline\hline
45.22148      &  $v$=1, 4(1,4)$-$3(1,3) E  & 193   \\
45.39579       & $v$=0, 4(1,4)$-$3(1,3) E  &  6  \\
45.39738       & $v$=0, 4(1,4)$-$3(1,3) A  &  6  \\
45.75404       & $v$=0, 3(1,3)$-$2(0,2) E   & 4   \\
45.75870       & $v$=0, 3(1,3)$-$2(0,2) A   & 4  \\
45.84739       & $v$=0, 14(3,11)$-$14(3,12) E  & 70      \\
45.88793       & $v$=0, 14(3,11)$-$14(3,12) A & 70   \\
81.31421    & $v$=0, 16(3,13)$-$16(2,14) E  & 89   \\
81.36235    & $v$=0, 16(3,13)$-$16(2,14) A  & 89     \\
81.38058    & $v$=0, 3(2,1)$-$2(1,2) E & 6   \\
82.24298    & $v$=0, 7(1,7)$-$2(0,6) E &  16 \\
82.24447    & $v$=0, 7(1,7)$-$2(0,6) A  & 16     \\
82.52349    & $v$=0, 19(4,15)$-$19(3,16) E &  126  \\
82.56197    & $v$=0, 19(4,15)$-$19(3,16) A & 126  \\
83.60516    & $v$=0, 10(3,8)$-$10(2,9) E & 38   \\
83.63843    & $v$=0, 10(3,8)$-$10(2,9) A & 38   \\
84.22466    & $v$=0, 11(4,7)$-$11(3,8) E & 50   \\
84.23334    & $v$=0, 11(4,7)$-$11(3,8) A  & 50   \\
84.28311    & $v$=1, 7(2,6)$-$6(2,5) E & 206   \\
84.44917    & $v$=0, 7(2,6)$-$6(2,5) E & 19    \\
84.45475    & $v$=0, 7(2,6)$-$6(2,5) A & 19     \\
85.32703    & $v$=1, 7(4,4)$-$6(4,3) A  & 215     \\
85.37173    & $v$=1, 7(3,5)$-$6(3,4) A & 210     \\
85.50622    & $v$=1, 7(4,3)$-$6(4,2) E & 215     \\
85.55338    & $v$=1, 7(5,3)$-$6(5,2) E & 220     \\
85.63833    & $v$=0, 4(2,3)$-$3(1,2) E & 9    \\
85.74398    & $v$=1, 7(4,4)$-$6(4,3) E & 214   \\    
85.77340    & $v$=0, 21(5,16)$-$21(4,17) A & 156   \\
85.78067    & $v$=0, 20(5,15)$-$20(4,16) E & 143   \\
85.78534    & $v$=0, 20(5,15)$-$20(4,16) A & 143   \\
85.91921    & $v$=0,  7(6,1)$-$6(6,0) E & 40   \\
86.02112    & $v$=0,  7(5,2)$-$6(5,1) E &  33   \\
86.02772    & $v$=0,  7(5,3)$-$6(5,2) E & 33  \\
86.02944    & $v$=0,  7(5,3)$-$6(5,2) A & 33     \\
86.03019    & $v$=0,  7(5,2)$-$6(5,1) A & 33   \\
86.03401   & $v$=1, 7(3,4)$-$6(3,3) E & 210   \\
86.15508    & $v$=1, 7(3,4)$-$6(3,3) A & 210   \\
86.17271    & $v$=1, 7(3,5)$-$6(3,4) E & 209   \\
86.21006    & $v$=0, 7(4,4)$-$6(4,3) A & 27   \\
86.22365    & $v$=0, 7(4,3)$-$6(4,2) E & 27  \\
86.22416    & $v$=0, 7(4,4)$-$6(4,3) E  & 27  \\
86.25055    & $v$=0, 7(4,3)$-$6(4,2) A & 27  \\
86.26580    & $v$=0, 7(3,5)$-$6(3,4) A & 23  \\
86.26874    & $v$=0, 7(3,5)$-$6(3,4) E &  23 \\
87.14328    & $v$=0, 7(3,4)$-$6(3,3) E  & 23   \\
87.16129    & $v$=0, 7(3,4)$-$6(3,3) A  & 23  \\
87.76638    & $v$=0, 8(0,8)$-$7(1,7) E & 20   \\    
87.76904    & $v$=0, 8(0,8)$-$7(1,7) A & 20   \\    
88.05397    & $v$=0, 19(5,14)$-$19(4,15) E & 130  \\
88.05446    & $v$=0, 19(5,14)$-$19(4,15) A & 130   \\
88.17551    & $v$=0, 10(4,6)$-$10(3,7) E & 43   \\
88.18042    & $v$=0, 10(4,6)$-$10(3,7) A & 43   \\
88.22075    & $v$=1, 7(1,6)$-$6(1,5) E & 205  \\
88.35849    & $v$=0, 22(5,17)$-$22(4,18) A & 170   \\
88.68689    & $v$=0, 11(3,9)$-$11(2,10) E & 45  \\    
88.72327    & $v$=0, 11(3,9)$-$11(2,10) A & 45  \\
88.77087    & $v$=1, 8(1,8)$-$7(1,7) A  & 208   \\
88.85161    & $v$=0, 7(1,6)$-$6(1,5) A  & 18  \\
88.86241    & $v$=1, 8(1,8)$-$7(1,7) E & 207  \\
98.42421    & $v$=0, 8(5,3)$-$7(5,2) E  & 38  \\
98.43180    & $v$=0, 8(5,4)$-$7(5,3) E & 38   \\
\hline
\end{tabular}
\end{center}
{$^{b}$ Integrated intensities derived from the LTE fit.} \\
\label{table-MF-single-dish}
\end{table*}  

\addtocounter{table}{-1}
\begin{table*}
\caption{(Continued).}
\begin{center}
%\vspace{-4mm}
\begin{tabular}{c c c}
\hline
Frequency &  Transition &  $E_{\rm up}$  \\
(GHz) &    & (K) \\
\hline\hline
98.43276    & $v$=0, 8(5,4)$-$7(5,3) A & 38   \\
98.43580    & $v$=0, 8(5,3)$-$7(5,2) A & 38  \\
98.60686    & $v$=0, 8(3,6)$-$7(3,5) E & 27  \\
98.61116    & $v$=0, 8(3,6)$-$7(3,5) A & 27    \\
98.68242    & $v$=1, 8(3,6)$-$7(3,5) E & 214   \\
98.68261    & $v$=0, 8(4,5)$-$7(4,4) A & 32     \\
98.71200    & $v$=0, 8(4,5)$-$7(4,4) E & 32   \\
98.74791    & $v$=0, 8(4,4)$-$7(4,3) E & 32  \\
98.79229    & $v$=0, 8(4,4)$-$7(4,3) E & 32    \\
110.05033   & $v$=1, 9(6,4)$-$8(6,3) E & 237    \\
110.15365   & $v$=1, 10(1,10)$-$9(1,9) A & 218   \\
110.23871   & $v$=1, 10(1,10)$-$9(1,9) E & 217   \\
111.40841   & $v$=0, 9(4,5)$-$8(4,4) E & 37  \\
111.45330   & $v$=0, 9(4,5)$-$8(4,4) A &  37 \\
111.67413   & $v$=0, 9(1,8)$-$8(1,7) E & 28   \\
111.68219   & $v$=0, 9(1,8)$-$8(1,7) A & 28   \\
111.73400   & $v$=0, 10(1,10)$-$9(0,9) E & 30  \\
111.73531   & $v$=0, 10(1,10)$-$9(0,9) A & 30    \\
140.16666   & $v$=1, 11(2,9)$-$10(2,8) A & 231   \\
147.24796   & $v$=0, 18(6,13)$-$18(5,14) E & 125  \\
147.24796   & $v$=1, 12(5,7)$-$11(5,6) E &  250  \\
147.25078    & $v$=0, 12(11,1)$-$11(11,0) E & 126   \\
147.25568    & $v$=0, 12(11,1)$-$11(11,0) A & 126   \\
147.25568    & $v$=0, 12(11,2)$-$11(11,1) A & 126     \\
147.26531    & $v$=0, 12(11,2)$-$11(11,1) E & 126   \\
147.30479    & $v$=0, 19(6,14)$-$19(5,15) E & 137    \\
147.31057    & $v$=0, 12(10,2)$-$11(10,1) E & 112   \\
147.31775    & $v$=0, 12(10,2)$-$11(10,1) A  & 112   \\
147.31775    & $v$=0, 12(10,3)$-$11(10,2) A &  112  \\
147.32539    & $v$=0, 12(10,3)$-$11(10,2) E & 112   \\
147.33163    & $v$=0, 19(6,14)$-$19(5,15) A & 137    \\
147.39707    & $v$=0, 12(9,3)$-$11(9,2) E & 100   \\
147.40637    & $v$=0, 12(9,3)$-$11(9,2) A & 100     \\
147.40637    & $v$=0, 12(9,4)$-$11(9,3) A &  100  \\
147.41182    & $v$=0, 12(9,4)$-$11(9,3) E & 100  \\
147.52431    & $v$=0, 12(8,4)$-$11(8,3) E & 89     \\
147.53554    & $v$=0, 12(8,5)$-$11(8,4) A & 88     \\
147.53554    & $v$=0, 12(8,4)$-$11(8,3) A & 88    \\
147.53864    & $v$=0, 12(8,5)$-$11(8,4) E & 88   \\
147.53917    & $v$=0, 17(6,12)$-$17(5,13) E & 115     \\
154.98454    & $v$=0, 12(3,9)$-$11(3,8) E & 53   \\
155.00232    & $v$=0, 12(3,9)$-$11(3,8) A & 53  \\
161.41614    & $v$=0, 13(5,8)$-$12(5,7) E & 71     \\
161.45743    & $v$=0, 20(2,18)$-$20(1,19) A & 128  \\
161.45822    & $v$=0, 13(5,8)$-$12(5,7) A & 70   \\
168.49507    & $v$=0, 13(3,10)$-$12(3,9) E & 61  \\
168.51375    & $v$=0, 13(3,10)$-$12(3,9) A &  61  \\
168.91461    & $v$=0, 15(2,13)$-$14(3,12) E & 76   \\
168.93457    & $v$=0, 15(2,13)$-$14(3,12) A & 76    \\
169.57068    & $v$=0, 20(7,13)$-$20(6,14) A & 157    \\
170.23327    & $v$=0, 14(3,12)$-$13(3,11) E & 68   \\
170.59390    & $v$=0, 6(4,2)$-$5(3,3) A & 23     \\
171.06714    & $v$=0, 23(7,17)$-$23(6,18) A & 196    \\
171.79421    & $v$=0, 14(13,1)$-$13(13,0) E & 174   \\
171.79451    & $v$=0, 14(13,1)$-$13(13,0) A  & 174   \\
171.79451    & $v$=0, 14(13,2)$-$13(13,1) A & 174   \\
171.84471    & $v$=0, 14(12,2)$-$13(12,1) E & 157   \\
171.84664    & $v$=0, 19(1,18)$-$19(1,19) E & 109   \\
171.84789    & $v$=0, 14(12,3)$-$13(12,2) A & 157    \\
171.84792    & $v$=0, 14(12,2)$-$13(12,1) A &  157  \\
171.84992    & $v$=0, 19(1,18)$-$19(0,19) E & 109   \\
171.86084    & $v$=0, 14(12,3)$-$13(12,2) E & 157    \\
171.88085    & $v$=0, 25(5,21)$-$25(4,22) E & 210     \\
171.91579    & $v$=0, 14(11,3)$-$13(11,2) E & 142  \\
\hline
\end{tabular}
\end{center}
%\label{table-MF-single-dish}
\end{table*}  

\addtocounter{table}{-1}
\begin{table*}
\caption{(Continued).}
\begin{center}
%\vspace{-4mm}
\begin{tabular}{c c c }
\hline
Frequency &  Transition &  $E_{\rm up}$ \\
(GHz) &    & (K) \\
\hline\hline
171.92174    & $v$=0, 14(11,3)$-$13(11,2) A &  142  \\
171.92174    & $v$=0, 14(11,4)$-$13(11,3) A &  142    \\
171.93258    & $v$=0, 14(11,4)$-$13(11,3) E & 142   \\
171.93339    & $v$=0, 19(1,18)$-$19(1,19) A & 109    \\
171.93528    & $v$=0, 25(5,21)$-$25(4,22) A & 210   \\
171.93658    & $v$=0, 19(1,18)$-$19(0,19) A  &  109  \\
171.99364    & $v$=0, 19(2,18)$-$19(0,19) E & 109   \\
172.01564    & $v$=0, 14(10,4)$-$13(10,3) E & 128   \\
172.02428    & $v$=0, 14(10,4)$-$13(10,3) A & 128   \\
172.02428    & $v$=0, 14(10,5)$-$13(10,4) A & 128     \\
172.15764    & $v$=0, 14(9,5)$-$13(9,4) E & 116   \\
172.16873    & $v$=0, 14(9,6)$-$13(9,5) A  &  116   \\
172.16873    & $v$=0, 14(9,5)$-$13(9,4) A & 116    \\
172.36437    & $v$=0, 14(8,6)$-$13(8,5) E &  104    \\
172.37775    & $v$=0, 14(8,7)$-$13(8,6) A & 104    \\
172.37775    & $v$=0, 14(8,6)$-$13(8,5) A & 104     \\
172.38095    & $v$=0, 14(8,7)$-$13(8,6) E & 104   \\
172.38578    & $v$=0, 15(1,14)$-$14(2,13) E &  71  \\
172.39349    & $v$=0, 15(1,14)$-$14(2,13) A &  71    \\
172.69216    & $v$=0, 14(7,8)$-$13(7,7) A & 95  \\
172.69337    & $v$=0, 14(7,8)$-$13(7,7) E & 95  \\
172.69337    & $v$=0, 14(7,7)$-$13(7,6) A & 95   \\
173.19427    & $v$=0, 14(6,9)$-$13(6,8) E &  86  \\
173.21868    & $v$=0, 14(6,8)$-$13(6,7) A & 86  \\
173.51541    & $v$=0, 15(2,14)$-$14(2,13) E &  71  \\
173.52168    & $v$=0, 15(2,14)$-$14(2,13) A & 71   \\
173.65011    & $v$=0, 14(4,11)$-$13(4,10) A & 73  \\
173.81936    & $v$=0, 14(5,10)$-$13(5,9) A & 79  \\
173.82245    & $v$=0, 14(5,10)$-$13(5,9) E & 79   \\
174.20980    & $v$=0, 15(1,14)$-$14(1,13) E & 71   \\
174.21556    & $v$=0, 15(1,14)$-$14(1,13) A & 71  \\
174.21802    & $v$=0, 27(6,22)$-$27(5,23) A & 249  \\
174.37741    & $v$=0, 14(5,9)$-$13(5,8) E &  79 \\
174.40618    & $v$=0, 14(5,9)$-$13(5,8) A & 79  \\
174.54656    & $v$=0, 16(0,16)$-$15(1,15) E & 73  \\
174.54801    & $v$=0, 16(0,16)$-$15(1,15) A & 73  \\
174.56580    & $v$=0, 16(1,16)$-$15(1,15) E & 73   \\
174.56721    & $v$=0, 16(1,16)$-$15(1,15) A & 73   \\
174.58112    & $v$=0, 16(0,16)$-$15(0,15) E  &  73  \\
174.58251    & $v$=0, 16(0,16)$-$15(0,15) A &  73  \\
174.60035    & $v$=0, 16(1,16)$-$15(0,15) E & 73  \\
174.60170    & $v$=0, 16(1,16)$-$15(0,15) A  & 73  \\
224.31315    & $v$=0, 18(5,14)$-$17(5,13) E  & 118  \\
224.60938    & $v$=0, 18(6,12)$-$17(6,11) A & 125   \\
237.29691    & $v$=0, 13(4,10)$-$12(3,9) E &  65 \\
237.29748    & $v$=0, 20(2,18)$-$19(2,17) E & 128   \\
237.30597    & $v$=0, 20(2,18)$-$19(2,17) A &  128   \\
237.30954    & $v$=0, 21(2,20)$-$20(2,19) E & 132  \\
237.31508    & $v$=0, 21(2,20)$-$20(2,19) A  & 132  \\
237.31705    & $v$=0, 13(4,10)$-$12(3,9) A  & 65  \\
237.34487    & $v$=0, 21(1,20)$-$20(1,19) E  & 132  \\
237.35039    & $v$=0, 21(1,20)$-$20(1,19) A  & 132 \\
258.08104    & $v$=0, 22(2,20)$-$21(2,19) E  & 152  \\
258.08949    & $v$=0, 22(2,20)$-$21(2,19) A  & 152 \\
258.30628    & $v$=0, 11(5,7)$-$10(4,6) A & 56   \\
258.49087    & $v$=0, 23(2,22)$-$22(2,21) E  &  156  \\
258.49624    & $v$=0, 23(2,22)$-$22(2,21) A  & 156  \\
258.49933    & $v$=0, 21(12,10)$-$20(12,9) E  &  232  \\
258.50273    & $v$=0, 23(1,22)$-$22(1,21) E  & 156   \\
258.50818    & $v$=0, 23(1,22)$-$22(1,21) A & 156  \\
\hline
\end{tabular}
\end{center}
%\label{table-MF-single-dish}
%\end{scriptsize}
\end{table*}

%-----------------------------Table Start-----------------------------

\begin{table*}
%\begin{scriptsize}
\caption[]{Unblended transitions (i.e. non blended with other molecular species) of MF identified in the interferometric SMA spectra towards G31.}
\begin{center}
%\vspace{-4mm}
\begin{tabular}{c c c }
\hline
Frequency &  Transition &  $E_{\rm up}$     \\
(GHz) &   & (K)\\
\hline\hline
219.62269     & $v$=1, 18(12,6)$-$17(12,5) A  & 384   \\
219.62269     & $v$=1, 18(12,7)$-$17(12,6) A  & 384   \\ 
219.64240     & $v$=1, 18(13,6)$-$17(13,5) E  & 401   \\
219.69583     & $v$=1, 18(11,8)$-$17(11,7) A  & 369   \\
219.69583     & $v$=1, 18(11,7)$-$17(11,6) A  & 369   \\
219.70513     & $v$=1, 18(4,15)$-$17(4,14) A  & 299   \\  
220.03034     & $v$=1, 18(9,10)$-$17(9,9) A   & 342   \\
220.03034     & $v$=1, 18(9,9)$-$17(9,8) A    & 342   \\
220.16689     & $v$=0, 17(4,13)$-$16(4,12) E  & 103   \\
220.25866     & $v$=1, 18(8,10)$-$17(8,9) E   & 331   \\
220.30738     & $v$=1, 18(10,9)$-$17(10,8) E  & 354   \\
220.36833     & $v$=1, 18(8,11)$-$17(8,10) A  & 331   \\
220.36988     & $v$=1, 18(8,10)$-$17(8,9) A   & 331   \\
220.91395     & $v$=1, 18(7,12)$-$17(7,11) A  & 321   \\
220.92617     & $v$=0, 18(16,2)$-$17(16,1) A  & 271   \\
220.92617     & $v$=1, 18(16,3)$-$17(16,2) A  & 271   \\
220.93518     & $v$=0, 18(16,2)$-$17(16,1) E  & 271  \\
220.94742     & $v$=0, 18(16,3)$-$17(16,2) E  & 271   \\
220.97782     & $v$=1, 18(15,4)$-$17(15,3) A  & 250   \\
220.97782     & $v$=1, 18(15,3)$-$17(15,2) A  & 250   \\
220.98367     & $v$=0, 18(15,3)$-$17(15,2) E  & 250   \\
220.98533     & $v$=1, 18(7,11)$-$17(7,10) E  & 321   \\
221.04766     & $v$=1, 18(14,5)$-$17(14,4) A  & 231   \\
221.04766     & $v$=1, 18(14,4)$-$17(14,3) A  & 231   \\
221.04999     & $v$=0, 18(14,4)$-$17(14,3) E  & 231   \\
221.06693     & $v$=0, 18(14,5)$-$17(14,4) E  & 231   \\
221.11067     & $v$=1, 18(8,11)$-$17(8,10) E  & 330   \\
221.13972     & $v$=0, 18(13,5)$-$17(13,4) E  & 213   \\
221.14113     & $v$=0, 18(13,5)$-$17(13,4) A  & 213   \\
221.14113     & $v$=0, 18(14,4)$-$17(14,3) A  & 213   \\
221.15854     & $v$=0, 18(13,6)$-$17(13,5) E  & 213   \\
229.40502     & $v$=0, 18(3,15)$-$17(3,14) E  & 111   \\
229.42034     & $v$=0, 18(3,15)$-$17(3,14) A  & 111   \\
230.87881     & $v$=1, 18(4,14)$-$17(4,13) A  & 301   \\
231.24542     & $v$=1, 19(4,16)$-$18(4,15) A  &  310  \\
\hline
\end{tabular}
\end{center}
\label{table-MF-SMA}
%\end{scriptsize}
\end{table*}

%%%%%%%%% GA

 %-----------------------------Table Start-----------------------------

\begin{table*}
%\begin{scriptsize}
\caption[]{Unblended transitions (i.e. non blended with other molecular species) of GA identified in the single IRAM 30m single dish spectra towards G31.}
\begin{center}
%\vspace{-4mm}
\begin{tabular}{c c c }
\hline

Frequency &  Transition &  $E_{\rm up}$   \\
(GHz) &   & (K) \\
\hline\hline
82.47067      & 8(0,8)$-$7(1,7)  & 19   \\
88.69126      & 12(3,10)$-$12(2,11)  & 49   \\
\hline
\end{tabular}
\end{center}
\label{table-GA-single-dish}
%\end{scriptsize}
\end{table*}  

%-----------------------------Table Start-----------------------------

\begin{table*}
%\begin{scriptsize}
\caption[]{Unblended transition of GA identified in the interferometric SMA spectra towards G31.}
\begin{center}
%\vspace{-4mm}
\begin{tabular}{c c c }
\hline
Frequency &  Transition &  $E_{\rm up}$   \\
(GHz) &    & (K) \\
\hline\hline
230.89847     & 21(2,19)$-$20(3,18)   & 131 \\
\hline
\end{tabular}
\end{center}
\label{table-GA-SMA}
%\end{scriptsize}
\end{table*}  

%%%%%%%%% DME

 %-----------------------------Table Start-----------------------------

\begin{table*}
%\begin{scriptsize}
\caption[]{Unblended transitions of DME identified in the IRAM 30m single dish spectra towards G31.}
\begin{center}
%\vspace{-4mm}
\begin{tabular}{c c c}
\hline
Frequency &  Transition &  $E_{\rm up}$   \\
(GHz) &    & (K) \\
\hline\hline
82.45696       & 11(1,10)$-$11(0,11) AE  &  63  \\
82.45696       & 11(1,10)$-$11(0,11) EA  &  63  \\
82.45862       & 11(1,10)$-$11(0,11) EE  &  63  \\
82.46038       & 11(1,10)$-$11(0,11) AA  &  63  \\
%\hline
82.64930       & 3(1,3)$-$2(0,2) EA  & 7   \\
82.64930       & 3(1,3)$-$2(0,2) AE & 7   \\
82.65018       & 3(1,3)$-$2(0,2) EE & 7  \\
82.65108       & 3(1,3)$-$2(0,2) AA & 7   \\
%\hline
82.68650       & 4(2,2)$-$4(1,3) AE  & 15   \\
82.68650       & 4(2,2)$-$4(1,3) EA  & 15   \\
82.68877       & 4(2,2)$-$4(1,3) EE  & 15  \\
82.69114       & 4(2,2)$-$4(1,3) AA  & 15  \\
%\hline
83.09738       & 14(2,12)$-$14(1,13) EA & 103  \\
83.09738       & 14(2,12)$-$14(1,13) AE & 103   \\
83.09892       & 14(2,12)$-$14(1,13) EE & 103   \\
83.10044       & 14(2,12)$-$14(1,13) AA & 103  \\
%\hline
84.63202       & 3(2,1)$-$3(1,2) AE & 11   \\
84.63202       & 3(2,1)$-$3(1,2) EA & 11  \\
84.63440       & 3(2,1)$-$3(1,2) EE & 11  \\
84.63680       & 3(2,1)$-$3(1,2) AA & 11  \\
%\hline
85.97322       & 13(2,12)$-$12(3,9) AA  & 88   \\
85.97610       & 13(2,12)$-$12(3,9) EE  & 88    \\
85.97894       & 13(2,12)$-$12(3,9) EA  & 88  \\
85.97894       & 13(2,12)$-$12(3,9) AE  & 88  \\
%\hline
88.70622       & 15(2,13)$-$15(1,14) EA  & 117   \\
88.70622       & 15(2,13)$-$15(1,14) AE  & 117   \\
88.70764       & 15(2,13)$-$15(1,14) EE  & 117  \\
88.70907       & 15(2,13)$-$15(1,14) AA  & 117  \\
%\hline
109.57140       & 8(2,7)$-$8(1,8) EA  & 38   \\
109.57140       & 8(2,7)$-$8(1,8) AE  & 38   \\
109.57409       & 8(2,7)$-$8(1,8) EE  & 38   \\
109.57678       & 8(2,7)$-$8(1,8) AA  & 38   \\
%\hline
111.74135       & 19(3,16)$-$19(2,17) AE & 187   \\
111.74135       & 19(3,16)$-$19(2,17) EA  & 187  \\
111.74279       & 19(3,16)$-$19(2,17) EE  & 187  \\
111.74424       & 19(3,16)$-$19(2,17) AA   & 187  \\
%\hline
111.78256       & 7(0,7)$-$6(1,6) AA & 25   \\
111.78365       & 7(0,7)$-$6(1,6) EE & 25   \\
111.78365       & 7(0,7)$-$6(1,6) AE & 25   \\
111.78403       & 7(0,7)$-$6(1,6) EA & 25   \\       
%\hline
111.81205       &  18(3,15)$-$18(2,16) AE & 170  \\       
111.81205       &  18(3,15)$-$18(2,16) EA & 170  \\       
111.81367       &  18(3,15)$-$18(2,16) EE & 170   \\       
111.81529       &  18(3,15)$-$18(2,16) AA & 170   \\     
%\hline
147.20209       &  6(3,4)$-$6(2,5) EA  &  32  \\
147.20376       &  6(3,4)$-$6(2,5) AE &  32  \\
147.20681       &  6(3,4)$-$6(2,5) EE &  32  \\
147.21071       &  6(3,4)$-$6(2,5) AA &  32   \\
%\hline
154.45366       & 11(1,10)$-$10(2,9) AA  & 63   \\
154.45508       & 11(1,10)$-$10(2,9) EE  & 63  \\
154.45650       & 11(1,10)$-$10(2,9) AE  & 63   \\
154.45651       & 11(1,10)$-$10(2,9) EA  & 63   \\
%\hline
169.74067       & 16(3,14)$-$16(2,15) EA & 137  \\
169.74067       & 16(3,14)$-$16(2,15) AE & 137  \\
169.74349       & 16(3,14)$-$16(2,15) EE & 137   \\
169.74631       & 16(3,14)$-$16(2,15) AA & 137   \\
%\hline
169.90032       & 16(2,15)$-$16(1,16) EA  &  128  \\
169.90032       & 16(2,15)$-$16(1,16) AE  &  128 \\
169.90369       & 16(2,15)$-$16(1,16) EE  &  128 \\
169.90705       & 16(2,15)$-$16(1,16) AA  &  128   \\
\hline
\end{tabular}
\end{center}
\label{table-DME-single-dish}
\end{table*}

\addtocounter{table}{-1}
\begin{table*}
\caption{(Continued).}
\begin{center}
%\vspace{-4mm}
\begin{tabular}{c c c}
\hline

Frequency &  Transition &  $E_{\rm up}$  \\
(GHz) &   & (K) \\
\hline\hline
173.29268       & 10(0,10)$-$9(1,9) AA & 49  \\
173.29297       & 10(0,10)$-$9(1,9) EE & 49  \\
173.29327       & 10(0,10)$-$9(1,9) AE & 49  \\
173.29327       & 10(0,10)$-$9(1,9) EA & 49   \\
\hline                     
174.89053       & 17(3,15)$-$17(2,16) EA  &  152  \\
174.89053       & 17(3,15)$-$17(2,16) AE  &  152  \\
174.89329       & 17(3,15)$-$17(2,16) EE  &  152  \\
174.89605       & 17(3,15)$-$17(2,16) AA  &  152  \\
\hline
209.51519       & 11(1,11)$-$10(0,10) EA  & 59 \\
209.51519       & 11(1,11)$-$10(0,10) AE  & 59 \\
209.51561       & 11(1,11)$-$10(0,10) EE  & 59  \\
209.51603       & 11(1,11)$-$10(0,10) AA  & 59  \\
\hline      
237.61880       & 9(2,8)$-$8(1,7) EA  & 46   \\                                                               
237.61881       & 9(2,8)$-$8(1,7) AE  & 46   \\                                                               
237.62089       & 9(2,8)$-$8(1,7) EE  & 46   \\
237.62297       & 9(2,8)$-$8(1,7) AA  & 46  \\       

\hline
\end{tabular}
\end{center}
%\label{table-DME-single-dish}
\end{table*}

  %-----------------------------Table Start-----------------------------

\begin{table*}
%\begin{scriptsize}
\caption[]{Unblended transitions (i.e. non blended with other molecular species) of DME identified in the interferometric SMA spectra towards G31.}
\begin{center}
%\vspace{-4mm}
\begin{tabular}{c c c }
\hline
Frequency &  Transition &  $E_{\rm up}$   \\
(GHz) &   & (K) \\
\hline\hline
220.84652     & 24(4,20)$-$23(5,19) AA  &  298  \\
220.84764     & 24(4,20)$-$23(5,19) EE  &  298  \\
220.84876     & 24(4,20)$-$23(5,19) AE  &  298  \\
220.84876     & 24(4,20)$-$23(5,19) EA  &  298  \\
220.89182     & 23(4,20)$-$23(3,21) EA  &  274  \\
220.89182     & 23(4,20)$-$23(3,21) AE  &  274 \\
220.89311     & 23(4,20)$-$23(3,21) EE  &  274  \\
220.89500     & 23(4,20)$-$23(3,21) AA  &  274 \\
221.19722     & 27(5,22)$-$27(4,23) AA  &  381  \\
221.19782     & 27(5,22)$-$27(4,23) EE  &  381  \\
221.19842     & 27(5,22)$-$27(4,23) AE  &  381  \\
221.19842     & 27(5,22)$-$27(4,23) EA  &  381  \\
230.14003     & 25(4,22)$-$25(3,22) EA  &  319  \\                     
230.14003     & 25(4,22)$-$25(3,22) AE  &  319 \\    
230.14137     & 25(4,22)$-$25(3,22) EE  &  319 \\
230.14271     & 25(4,22)$-$25(3,22) AA  &  319  \\
230.23215     & 17(2,15)$-$16(3,14) AA  &  148  \\
230.23376     & 17(2,15)$-$16(3,14) EE  &  148  \\
230.23536     & 17(2,15)$-$16(3,14) AE  &  148  \\
230.23536     & 17(2,15)$-$16(3,14) EA  &  148  \\
\hline
\end{tabular}
\end{center}
\label{table-DME-SMA}
%\end{scriptsize}
\end{table*}

\clearpage

%%%%%%%%%%%%%%%%%%%%%%%%%%%%%%%%%%%%%%%%%%%%%%%

%%%%%%%%%%%%%%%%%%%%%%%%%%%%%%%%%%%%%%%%%%%%%%%

\section{MADCUBAIJ LTE fits of the GBT and SMA data}
\label{MADCUBA-fits}

In this Appendix we present the physical parameters used to model with {\it MADCUBAIJ} the transitions of the different COMs in the GBT and SMA spectra (Figs. \ref{fig-COMs-GBT} and \ref{figure-SMA-spectra}, respectively). The procedure used is: 
i) we have fixed the velocity of the emission and the linewidth to the values obtained from gaussian fits of unblended lines; 
ii) to obtain the source-average column density $N_{\rm s}$, we have fixed the size of the molecular emission to the value obtained from the integrated maps of unblended lines detected with the SMA, and applied the beam dilution factor accordingly; while we have not applied the beam dilution factor to derive the beam-average column density $N_{\rm b}$;
iii) leaving {\it N} and $T_{\rm ex}$ as free parameters, we have used the {\it AUTOFIT} tool of {\it MADCUBAIJ} to find the pair of values value that better fits the unblended molecular transitions of each COM.

%-----------------------------Table Start-----------------------------

\begin{table*}
%\begin{scriptsize}
\caption[]{MADCUBAIJ best fits of the COMs detected in the GBT spectra. The parameters of this table have been used to produce the synthetic spectra of Fig. \ref{fig-COMs-GBT}.}
\begin{center}
%\vspace{-4mm}
\begin{tabular}{c| c| c| c}
\hline
 & \multicolumn{1}{c|}{EG} & \multicolumn{1}{c|}{ET} & \multicolumn{1}{c}{MF}  \\
\hline
log$N_{\rm b}$ (cm$^{-2}$) & 15.1  & 15.5  & 16.1 \\
log$N_{\rm s}$ (cm$^{-2}$) & 17.9  & 17.7  & 18.1 \\
$T_{\rm ex}$ (K)$^{a}$ &  147$^{b}$ / 142$^{b}$ & 89 / 76   &  147 / 142  \\
$v_{\rm LSR}$ (km s$^{-1}$)  & 97.0 & 99.0  & 97.5  \\
$\Delta v$ (km s$^{-1}$) & 5.5 & 5.4    &  5.3 \\
%$\Delta E$ (K) & 5$-$9   & 70$-$88 & 4$-$193 \\
\hline
\end{tabular}
\end{center}
\begin{scriptsize}
\vspace{-2mm}
%\hspace{5cm}
%\centering
{${(a)}$ The first temperature corresponds to the beam-average value, while the second is the source-average value obtained assuming the source diameter of Table \ref{table-sizes}.} 
\\
{${(b)}$ Since the energy range covered by the transitions is very narrow, we have fixed the temperature to the values obtained for MF.} \\
\vspace{-4mm}
\label{table-physical-parameters-GBT}
\end{scriptsize}
\end{table*}

%-----------------------------Table Start-----------------------------

\begin{table*}
%\begin{scriptsize}
\caption[]{MADCUBAIJ best fits of the SMA compact spectra towards the peak of the continuum for the different molecules. The parameters of this table have been used to produce the synthetic spectra of Fig. \ref{figure-SMA-spectra}.}
\begin{center}
%\vspace{-4mm}
\begin{tabular}{c| c| c| c| c| c }
\hline
 & \multicolumn{1}{c|}{EG} & \multicolumn{1}{c|}{ET} & \multicolumn{1}{c|}{GA}   & \multicolumn{1}{c|}{MF} &   \multicolumn{1}{c}{DME}  \\
\hline
log$N_{\rm b}$ (cm$^{-2}$) & 16.3  & 17.2   & 16.0 & 17.5 & 17.7 \\
log$N_{\rm s}$ (cm$^{-2}$) & 17.7  &  17.7  & 17.4 & 17.9 & 18.7   \\
$T_{\rm ex}$ (K)$^{a}$ & 88 / 75 & 187 / 184 & 165$^{a}$  & 240 / 243  & 113 / 105\\
$v_{\rm LSR}$ (km s$^{-1}$) & 97.2  & 97.0  & 97.0  & 97.0  &  97.2 \\
$\Delta v$ (km s$^{-1}$) &  5.2  & 5.9 & 6.0 & 6.4 & 7.6 \\
%$\Delta E$ (K) &  137$-$160  & 85$-$263 & 131 & 103$-$400 & 147$-$380 \\
\hline
\end{tabular}
\end{center}
\begin{scriptsize}
\vspace{-2mm}
%\hspace{5cm}
\centering
{${(a)}$ The first temperature corresponds to the beam-average value, while the second is the source-average value obtained assuming the sourze diameter of Table \ref{table-sizes}.} 
\\
\vspace{-4mm}
\label{table-physical-parameters-SMA}
\end{scriptsize}
\end{table*}

\end{document}